\documentclass[lettersize, journal]{IEEEtran}
\usepackage[english]{babel}
\usepackage{blindtext}
\usepackage{subcaption}
\usepackage{hyperref}
\usepackage{enumitem}
\usepackage{multirow}

\usepackage{cite}
\usepackage{amsmath,amssymb,amsfonts}
\usepackage{algorithmic}
\usepackage{graphicx}
\usepackage{caption}
\usepackage{booktabs}
\usepackage{subcaption}
\usepackage{textcomp}
\usepackage{xcolor}
\usepackage[T1]{fontenc}
\usepackage{lmodern}
\usepackage[export]{adjustbox}
\usepackage{enumitem}

\ifCLASSINFOpdf
\else

\fi

\begin{document}
\title{Handover Configurations in Operational 5G Networks: Diversity, Evolution, and Impact on Performance}

\author{Moinak Ghoshal,~\IEEEmembership{Student Member,~IEEE,} Imran Khan, Phuc Dinh,~\IEEEmembership{Student Member,~IEEE,} Z. Jonny Kong, Omar Basit, Sizhe Wang, Yufei Feng, Y. Charlie Hu,~\IEEEmembership{Fellow,~IEEE}, and Dimitrios~Koutsonikolas,~\IEEEmembership{Senior Member,~IEEE}

\thanks{Moinak Ghoshal, Imran Khan, Phuc Dinh, Sizhe Wang, Yufei Feng, and Dimitrios
Koutsonikolas are with the Department of Electrical and Computer
Engineering, Northeastern University, Boston, MA 02115 USA
(e-mail: ghoshal.m@northeastern.edu; khan.i@northeastern.edu; dinh.p@northeastern.edu; wang.sizh@northeastern.edu; feng.yuf@
northeastern.edu; d.koutsonikolas@northeastern.edu).}
\thanks{Z. Jonny Kong, Omar Basit, and Y. Charlie Hu are with the Elmore Family School of Electrical and Computer
Engineering, Purdue University, EE Building, 465 Northwestern Ave., West Lafayette, IN 47907-2035
(e-mail:  kong102@purdue.edu; obasit@purdue.edu; ychu@purdue.edu).}
}
\maketitle
\IEEEpeerreviewmaketitle
\begin{abstract}

  Mobility management in cellular networks, especially the handover (HO) process,
plays a key role in providing seamless and ubiquitous
Internet access.
The wide-scale deployment of 5G and the resulting
co-existence of 4G/5G in the past six years
have significantly changed the landscape of all mobile network operators
and made the HO process much more complex than before.
While several recent works have studied the impact of HOs
on user experience,
why and how HOs occur and how HO configurations affect performance in 5G operational networks 
remains largely unknown.
%
Through four cross-country driving trips across the US spread out over a 27-month period, we conduct an
in-depth measurement study of HO configurations across all three
major US operators.
Our study 
reveals (a)
new types of HOs and new HO events 
used by operators to handle these new types of HOs, (b) overly aggressive HO configurations that result in unnecessarily high signaling overhead,
(c) large diversity in HO configuration parameter values, which also differ across operators, 
but significantly lower diversity in 5G compared to LTE, and (d) sub-optimal
HO configurations/decisions leading to poor pre- or post-HO
performance.
Our findings have many implications for mobile operators, as they keep fine-tuning their 5G HO configurations.
\end{abstract}

\vspace{-0.1in}
\section{Introduction}
\label{sec:intro}

Mobility management in cellular networks, and especially the handover
(HO) process, plays a key role in providing seamless and ubiquitous
internet access. Mobility management in operational cellular networks
follows a policy-based approach. Each HO takes into account many
factors, including cell priorities, signal strength, a list of events
of interest, candidate cells, etc. and is implemented via multiple
asynchronous procedures (measurement, reporting, decision and
execution), each with its own configuration parameters. While the
logic implementing these procedures is standardized in 3GPP
specifications, cellular operators are free to select different values
for each configurable parameter to implement different HO policies.

A previous study of these policy-based configurations focused on 4G
operational networks~\cite{deng:imc2018} and showed that 4G HOs are extremely complex in practice.
%
%
The wide-scale deployment of 5G in the past six years
has significantly changed the landscape of all mobile network operators.
%
5G's support for diverse radio bands (low, mid, high), different architectures 
(non-standalone (NSA) vs. standalone (SA)), and the coexistence of
different radio access technologies (RATs) (LTE, 5G) make mobility
management in the 5G era even more complex.
Such added complexity 
motivates the need for revisiting policy-based configurations in cellular network mobility management.

\begin{table}[tp]
    \centering
    \caption{Dataset statistics.}
    \label{tab:dataset_stats}
    \resizebox{1\columnwidth}{!}{
    \begin{tabular}{l|c}
    \toprule
    Measurement dates & August 2022 - November 2024 \\ \hline
    Total distance traveled  & 15000+ km \\ \hline
    States/major cities/counties  & 25/19/100+ \\ \hline
    Operators & Verizon (V), T-Mobile (T), AT\&T (A) \\ \hline
    \# of HOs & 18653 (V), 14091 (T), 15682 (A) \\ \hline
    \# of HOs by type & \begin{tabular}[c]{@{}c@{}}20139 (LTE$\rightarrow$ LTE), 8413 (LTE$\rightarrow$ 5G),\\ 3100 (5G$\rightarrow$ LTE), 16774 (5G$\rightarrow$ 5G)\end{tabular} \\ \hline
    \# of HO reporting events & 57948 (V), 72730 (T), 108485 (A) \\ \hline
    Total cellular data used & 9.3+ TB (Rx), 1.2+ TB (Tx) \\ \hline
    Total log size & 1.6+ TB \\ \hline
    Cumulative experiment runtime & 6400+ min per operator  \\ \bottomrule
    \end{tabular}}
   \vspace{-0.3in}
\end{table}

A number of recent measurement studies~\cite{narayanan:www2020,narayanan2021variegated,hassan:sigcomm2022,ghoshal:imc2023,xu:sigcomm:2020,khan:meditcom2024,kousias:Wintech2022,fiandrino:mswim2022} examined HOs in 5G operational networks and showed that they can result in large throughput fluctuations
and have a severe impact on the QoE of latency-critical 5G "killer apps". 
However, all these works focus on
{\em the outcome} of the HO process -- the impact on performance and user experience -- but ignore the {\em root cause}. To our best knowledge, no work has tried to shed light on (i) why and how HOs are triggered and (ii) how HO configurations affect performance in 5G operational networks. 

This work fills this gap by conducting the first large-scale, in terms of both geographic area and time,
study of HO configurations in the 5G era. Through 4 cross-country
driving trips (Los Angeles to Boston in Aug. '22 and Nov. '24,
Boston to Chicago in May '23, and Boston to Atlanta in Aug. '23)
covering a total of 15,000+ km and spread out over a 27-month period, we
collect a large dataset (Table~\ref{tab:dataset_stats})
of 48,426 HOs and 239,163 HO events spanning all 3 major US
operators (Verizon, AT\&T, T-Mobile), different RATs (5G vs. 4G), 5G
architectures (NSA vs. SA), and 5G bands (low, mid, high).



Using this rich dataset, we perform a detailed analysis of the HO landscape in the 5G era. 
We analyze different types of HOs and HO-triggering events, including several new types that did not exist in the pre-5G era, the configurations employed by different operators, the complexity of those configurations, their differences with respect to several dimensions (technology, frequency band, 5G architecture, area type), and their impact on performance. Additionally, the datasets from the two trips from Los Angeles to Boston, following the same route and conducted 27 months apart, allow us to analyze the evolution of the HO landscape and configurations over a two-year period of 5G deployment. 
The key findings of our study are summarized as follows:



\noindent $\bullet${\bf HO-triggering events.} The 5G HO landscape is very different from the LTE landscape, involving new HO-triggering events that support inter-RAT HOs, while some events present in LTE have been practically eliminated. 
While UEs report a very large number of LTE and 5G events, most of these events (69-87\% for different operators) do not trigger HOs, resulting in unnecessarily high overhead (\S\ref{sec:events}).


\noindent $\bullet${\bf Complexity.} Despite the increased complexity of the HO
landscape in the 5G era due to the existence of many different types
of HOs, operators tend to prefer one event for each HO type. However, we also discovered a
number of other reporting events appearing rarely,
including some unusual cases. While such rare occurrences may currently appear to be non-significant due to the limited number of HOs between certain bands and the limited use of the SA architecture, they raise concerns about increased complexity in the near future, when 5G deployments will have taken their final form (\S\ref{sec:breakdown}).


\noindent $\bullet${\bf Configuration diversity.} Our analysis of HO configurations reveals that most parameters exhibit high degrees of
diversity in terms of richness, distribution, and dispersion of
their values. Compared to 2018~\cite{deng:imc2018}, LTE configurations have become even more complex, with some parameters having tens of different values. 
5G configuration parameters, on the other hand, have a much smaller number of values (typically less than 10), but they still exhibit large diversity in terms of distribution and dispersion. Operators prefer to use same values for less popular events across technologies, 5G bands, and area types, but choose to fine-tune the configurations of popular events, although with some exceptions (\S\ref{sec:param}).


\noindent $\bullet${\bf Evolution.} Our comparison of the 2022 and 2024 datasets reveals that operators have stopped tweaking their LTE configurations, but continue fine-tuning their 5G configurations, 
although they prioritize different HO types. While new values for many parameters have been introduced over the past two years, operators are also converging towards a set of values for each parameter that are used across 5G bands and across different area types (\S\ref{sec:param}). We also found one event (A2) that has been practically eliminated in 2024 although it was responsible for 30-41\% of the 5G\textrightarrow LTE HOs in 2022 (\S\ref{sec:breakdown}). 


\noindent $\bullet${\bf Impact of configurations on performance.} The selection of optimal configurations remains very challenging for operators, as they try to strike a balance between better post-HO performance and timing (which can deteriorate pre-HO performance), while also mitigating the "ping pong" effect. 
Predicting the relative performance among different configurations can be challenging and optimizing all three metrics simultaneously is extremely tough, as pre-HO performance and ping pong HOs are also affected by other factors, such as the overall coverage. Additionally, the weak correlation between signal strength metrics, which operators have direct control over via their HO configurations, and throughput (the metric of interest from the user's perspective), makes it extremely tough to control the latter or predict its outcome based on the signal strength outcome (\S\ref{sec:throughput}).



\noindent\textbf{Contributions.} In summary, this work makes the
following contributions: (1) We collect a large multi-carrier,
multi-technology, multi-band dataset of cellular mobility
management during four cross-country driving trips spanning 15000+ km and spread out over a 27-month period.
%
(2)
We provide the first in-depth study of mobility
management in the 5G era \textit{focusing on the root cause (policy-based configurations and their impact on performance) and thus complementing previous studies, which focused on the outcome (impact of HOs on transport and application layer performance).}
(3) Our dataset will be open-sourced upon acceptance of the paper.


\section{HO Primer}
\label{sec:primer}

A HO is used to transfer a user's network session from the cell it is attached to (serving cell) to another cell (neighboring cell), because either the user moves out of the serving cell's coverage area or there is another cell operating at a different RAT. HOs can be classified into idle-state HOs and active-state HOs, based on whether the UE is idle or sending/receiving traffic to/from the network. Idle-state HOs are performed by the device, which selects an appropriate cell for future access. 
Active-state HOs are initiated by the network; the serving cell migrates the UE to another neighboring cell. In this work, we focus on active-state HOs.

\begin{figure}[t]
    \centering
    \includegraphics[width=0.8\columnwidth]{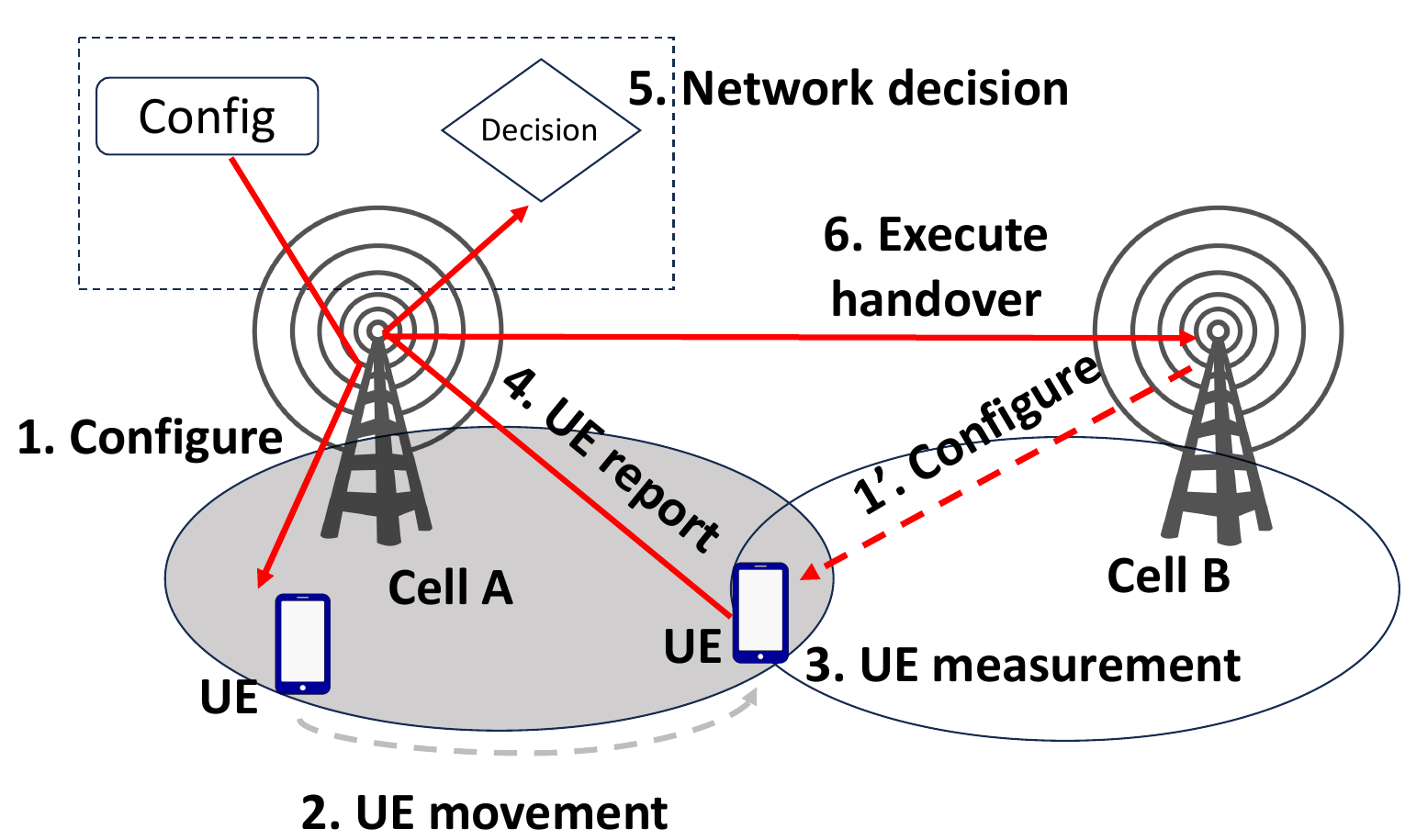}
    \vspace{-0.1in}
    \caption{Basic HO procedure.}
    \label{fig:ho_procedure}
    \vspace{-0.25in}
\end{figure}

Fig.~\ref{fig:ho_procedure} shows a basic cellular HO procedure. 
When a UE connects to a new cell, the cell advertises a HO configuration (1) to the UE. 
Based on the configuration, the UE learns the criteria to trigger a HO, including whether to invoke measurement, what and when to measure, and when/what to report. 
As the UE starts moving from the coverage area of Cell A to a new coverage area under Cell B, it starts measuring (3) different signal metrics (RSRP, RSRQ). 3GPP defines a set of measurement report (MR) mechanisms based on \textit{events}, which are triggered when certain signal-related criteria are met. When an event is triggered, the UE sends an MR (4), which the network uses to decide (5) whether to perform a HO or not. Based on the network's decision, the HO is executed (6) from Cell A to B, followed by a new set of configurations (1') advertised from Cell B to the UE.  

\begin{table}[t]
\centering
\caption{LTE/5G-NR HO-related reported events \footnotesize{($M$ = measurement, $C$ = serving cell, $N$ = neighboring cell)}} 
\label{tab:meas_events}
\resizebox{\columnwidth}{!}{
\begin{tabular}{|c|l|l|} 
\hline
\textbf{Event Type} & \textbf{Event Description}   &\textbf{trigger Condition}        \\ 
\hline
A1         & Serving cell becomes better than a threshold  & $M_S > \Theta_{A_1} + H_{A_1}$  \\ 
\hline
A2         & Serving cell becomes worse than a threshold  & $M_S < \Theta_{A_2} - H_{A_2}$    \\ 
\hline
A3         & Neighboring cell becomes offset better than the serving cell  & $M_N > M_S + \Delta_{A_3} + H_{A_3}$  \\ 
\hline
A4         & Neighboring cell becomes better than threshold     & $M_N > \Theta_{A_4} + H_{A_4}$  \\ 
\hline
A5         & \begin{tabular}[c]{@{}l@{}}Serving cell becomes worse than a threshold\\and neighboring cell becomes better than another threshold\end{tabular} & \begin{tabular}[c]{@{}l@{}}$ M_S < \Theta^S_{A_5} - H_{A_5}$ \\ $M_N > \Theta^N_{A_5} + H_{A_5}$ \end{tabular}\\ 
\hline
A6         & Neighboring cell becomes offset better than the serving cell  &$M_N > M_S + \Delta_{A_6} + H_{A_6}$ \\ 
\hline
B1         & Inter-RAT neighbor becomes better than threshold    & $M_N > \Theta_{B_1} + H_{B_1}$ \\ 
\hline
\begin{tabular}[c]{@{}c@{}}B2 \\ (5G only) \end{tabular}      & \begin{tabular}[c]{@{}l@{}}Primary cell becomes worse than a threshold \\and inter-RAT neighbor becomes better than a threshold\end{tabular} & \begin{tabular}[c]{@{}l@{}}$M_S < \Theta^S_{B_2} - H_{B_2}$ \\ $M_N > \Theta^N_{B_2} + H_{B_2}$\end{tabular}    \\
\hline 
P         &  \begin{tabular}[c]{@{}l@{}}Periodic reporting of serving and neighboring cell's radio \\signal quality.\end{tabular}    & N/A  \\
\hline
\end{tabular}}
\end{table}

\begin{table}[t]
\centering
\caption{Description of common event parameters.}
\label{tab:params}
\resizebox{\columnwidth}{!}{
\begin{tabular}{|l|l|} 
\hline
Parameter Name   & \textbf{Description}                                                                                                                                                                                                                                              \\ 
\hline
Hysteresis  $H$     & \begin{tabular}[c]{@{}l@{}}Prevents unnecessary and fast HOs. Sets a threshold below \\which the serving cell is considered better than a neighboring cell, \\even if the neighboring cell's signal strength is slightly stronger.\end{tabular}                                          \\ 
\hline
Threshold  $\Theta$      & \begin{tabular}[c]{@{}l@{}}Defines the level at which a measurement is triggered. When the \\measured value crosses the threshold, a measurement report \\is generated.\end{tabular}    \\ 
\hline
Offset  $\Delta$         & \begin{tabular}[c]{@{}l@{}}Added to the signal strength of a serving or a neighboring cell. \\Creates a margin in the comparison between the serving \\cell and the neighboring cell.\end{tabular}  \\
\hline
Time To Trigger   TTT      & \begin{tabular}[c]{@{}l@{}}Time during which specific criteria for the event need to be met \\in order to trigger a measurement report.\end{tabular}  
\\ 
\hline
Trigger Quantity & \begin{tabular}[c]{@{}l@{}}Defines the measured parameter (RSRP or RSRQ) used to trigger \\an event.\end{tabular}  \\
\hline
\end{tabular}}
\vspace{-0.2in}
\end{table}

Table~\ref{tab:meas_events} shows the list of HO related events prevalent in LTE/5G networks. Each event has its own configuration parameters, the most important of which are described in Table~\ref{tab:params}. While the events and parameters are standardized in 3GPP specifications, cellular operators are free to select different configurations (i.e., values for each parameter) implementing different HO policies.

\section{Methodology}
\label{sec:method}

\noindent\textbf{5G  Carriers.} We purchased multiple unlimited data plans from all three major US service providers, Verizon, T-Mobile, and AT\&T, to perform the experiments. All three carriers have 5G-low band, 5G-mid band, and 5G-high band (mmWave) services deployed all across the country, predominantly co-existing with LTE via Non-Stand Alone (NSA) architecture. T-Mobile was the only US carrier that supported both NSA and SA 5G services in the US at the time of the study.

\noindent\textbf{UEs and measurement tools.} We used Samsung S21 Ultra (in 2022, 2023) and S24 (in 2024) phones as UEs, which were the state-of-the-art devices at the time of our measurement campaigns. 
Since we only focus on \textit{active-state} HOs, which are controlled by the network, \textit{our findings are independent of the UE model and can be generalized for any smartphone model}. We performed a mix of backlogged downlink/uplink
throughput, ICMP-based ping RTT, and application tests in
a round-robin fashion throughout the driving trips. Each UE was connected to an Accuver XCAL Solo
device~\cite{accuverxcalsolo}, which taps into the Qualcomm diagnostic
(Diag) interface of the smartphone and extracts PHY layer KPIs and RRC layer signaling messages such as HOs, configurations, and MRs. 
We performed measurements \textit{concurrently} over the three operators, using 3 UES, each connected to a different operator.



\noindent\textbf{Driving Tests.} We drove 15500+ km during the four road trips covering all major cities on each route. In addition to driving on highways, we conducted additional downtown driving tests (one hour each) in select major cities on our routes to obtain data from urban environments. 

\noindent\textbf{Extracting HO-triggering events.}
The authors in~\cite{deng:imc2018} assumed that the
last reporting event is the one that triggers a HO. However, we found that this assumption is not
always valid in practice, especially
in NSA deployments. Additionally, we found several special cases, e.g., an NR SCG Modification, which is typically used to switch 5G cells~\cite{TS37340-multiconnectivity}, is sometimes used in LTE-5G HOs and its triggering MR is present in an LTE UL DCCH message instead of a 5G one. Therefore, we developed a new methodology, for extracting HO-triggering events and configurations. Our methodology addresses three challenges:

\begin{figure}[t]
        \includegraphics[width=\columnwidth]{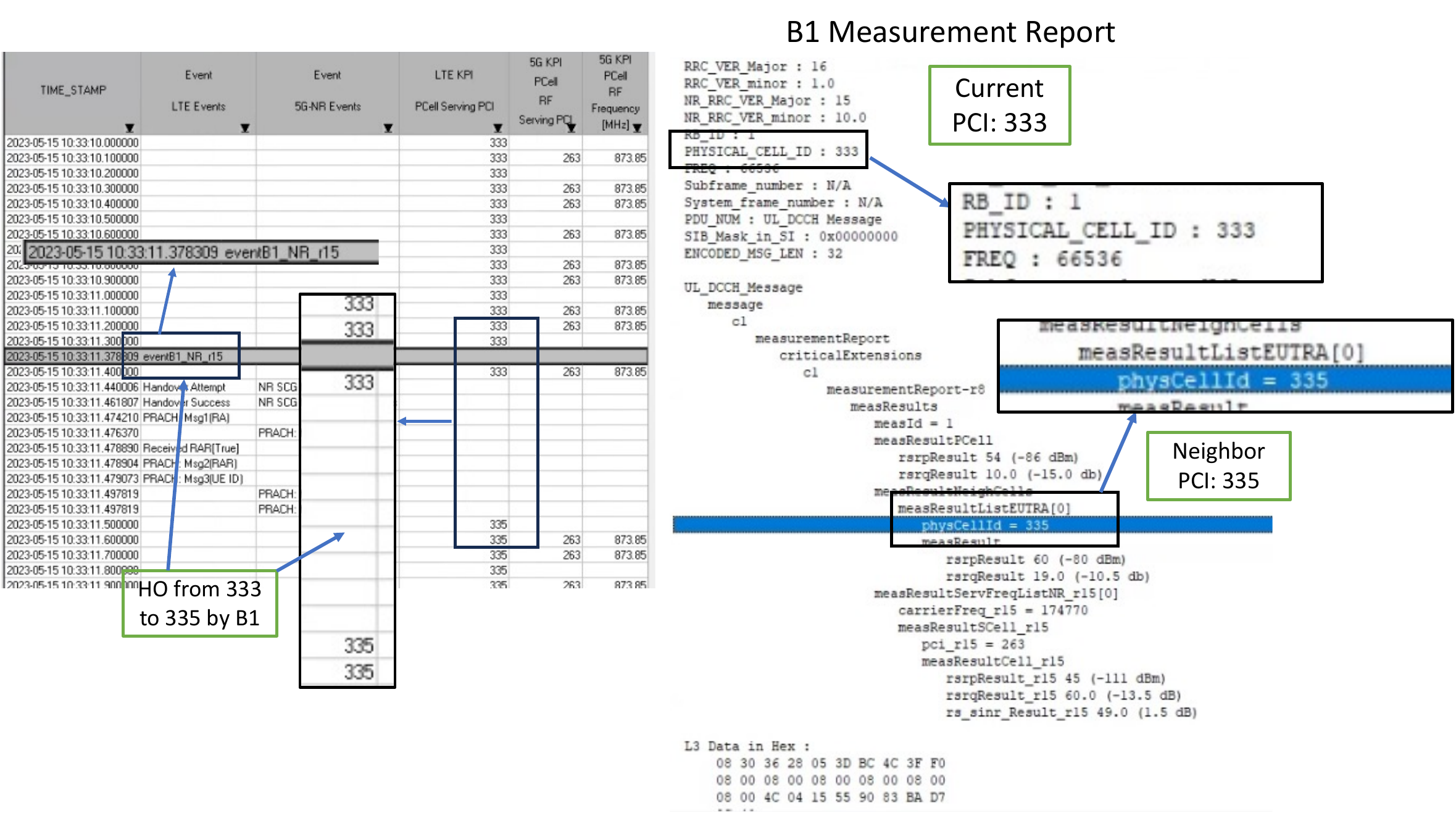}
    \caption{Example showing B1 report triggering an LTE-LTE HO.}
    \label{fig:b1_lte_lte}
\end{figure}

\begin{figure}[t]
        \includegraphics[width=\columnwidth]{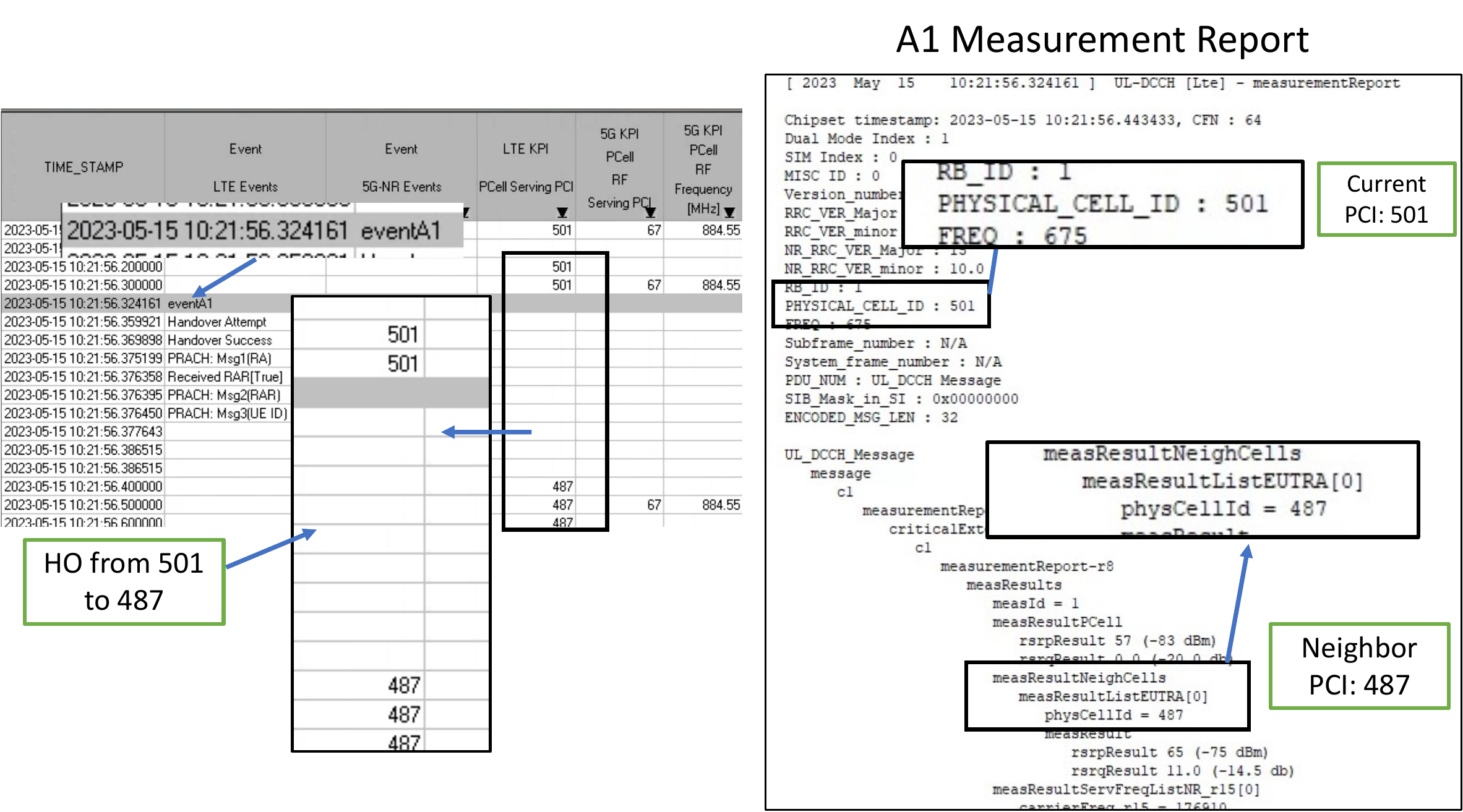}
    \caption{Example showing A1 report triggering an LTE-LTE HO.}
    \label{fig:a1_lte_lte}
\end{figure}

\begin{figure}[t]
        \includegraphics[width=\columnwidth]{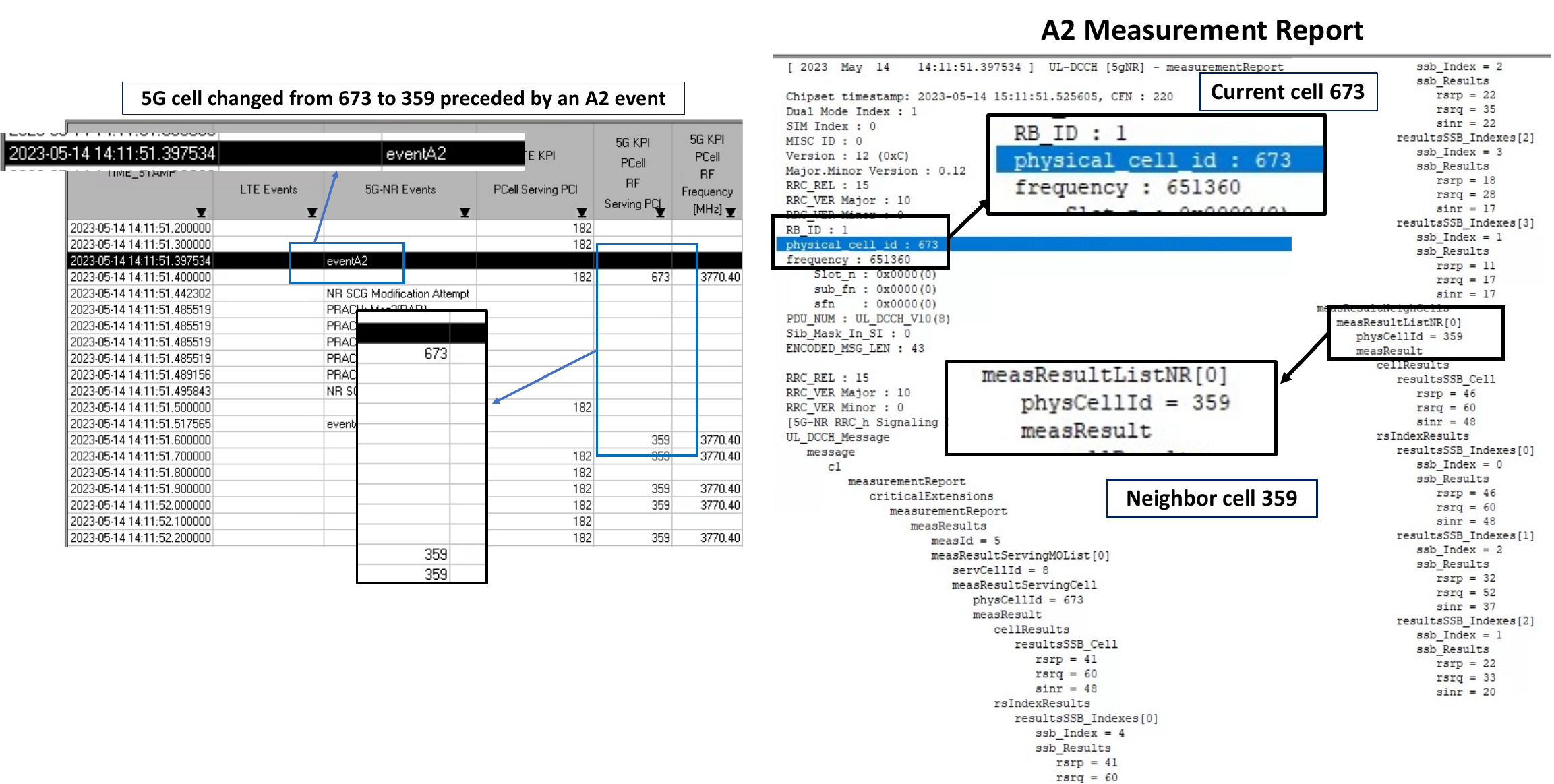}
    \caption{Example showing A2 report triggering an 5G-5G HO.}
    \label{fig:a2_5G_5G}
\end{figure}

\noindent $\bullet$ \textbf{C1} The authors in~\cite{deng:imc2018} analyzed LTE HOs assuming that the
last reporting event is the one that triggers a HO, as most of the HOs
happen within 80-230 ms after the last MR is sent to
the connected cell by the UE. This assumption is not
always valid in the 5G era, due to the increased number of different HO types
and the co-existence of LTE and 5G technologies, especially in NSA
deployments.

\noindent $\bullet$ \textbf{C2} In general, MRs pertaining to LTE HOs are sent via LTE UL DCCH messages and those pertaining to 5G via 5G UL DCCH messages. However, with NSA still being the dominant 5G architecture, we observed many exceptions in practice. For example, an NR SCG Modification is typically used to switch 5G cells~\cite{TS37340-multiconnectivity}, but we found cases where it is used in LTE-5G HOs and its triggering MR is present in an LTE UL DCCH message instead of a 5G one. 

\noindent $\bullet$ \textbf{C3} When the UE connects to a new cell after a HO, the new configuration parameters are advertised by the network via an RRC message. However, in the case of 5G,\footnote{We also found a few instances in LTE.} the network sends a partial version of the RRC message if the UE had previously connected to the current cell within a certain time.

To address challenges \textbf{C1} and \textbf{C2}, we inspect of all MRs preceding a HO, present in both LTE and 5G UL DCCH messages (for 5G NSA HOs) to find the one containing the the cell ID and frequency that match those of the new cell. \textit{This allowed us to correctly identify the correct triggering event for 12\% of the HOs in our dataset compared to the simplified methodology from~\cite{deng:imc2018}, and to uncover some unusual yet interesting cases}. For example, a B1 event, which is used for LTE to 5G handovers, can sometimes trigger a LTE to LTE handover (Fig.~\ref{fig:b1_lte_lte}). We also found cases where a HO was triggered by an A1 or A2 event. Typically, the UE uses A1 (A2) events to report that its signal strength became better (worse) than a threshold, which cancels
a HO or triggers another event (e.g., A3, A5, or B1, in the near future). However, we found certain A1/A2 MRs also report signal strength of neighboring cells, which makes a HO possible (Figs.~\ref{fig:a1_lte_lte},~\ref{fig:a2_5G_5G}). 

Finally, to address challenge \textbf{C3} for a certain number of HOs in our dataset, we have to trace back through multiple HOs to extract
complete configuration parameters of the current cell.

\section{Overview of HO-Related Reporting Events}
\label{sec:events}

\begin{figure}
    \begin{subfigure}[b]{0.32\linewidth}
        \centering
        \includegraphics[width=\linewidth]{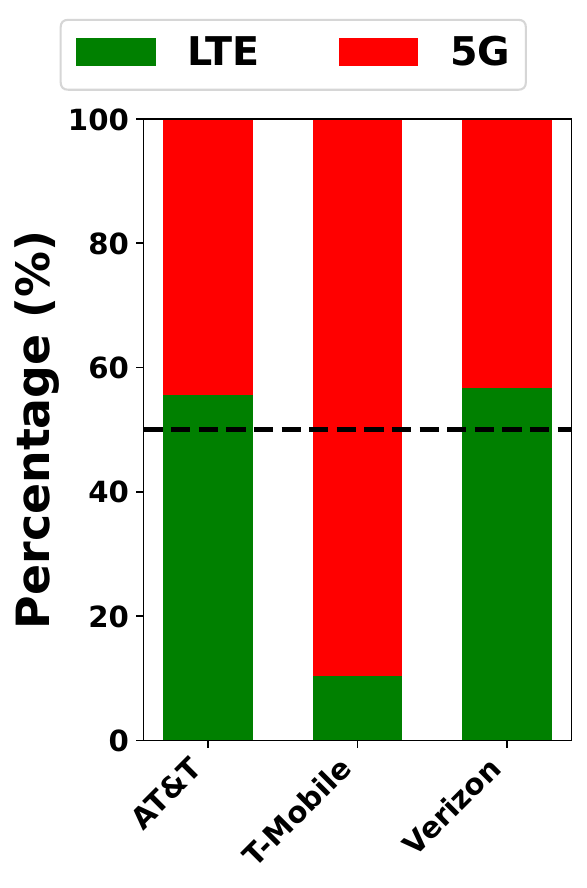}
        \caption{Overall.}
        \label{fig:coverage_overall}
    \end{subfigure}   
     \begin{subfigure}[b]{0.32\linewidth}
        \centering
        \includegraphics[width=\linewidth]{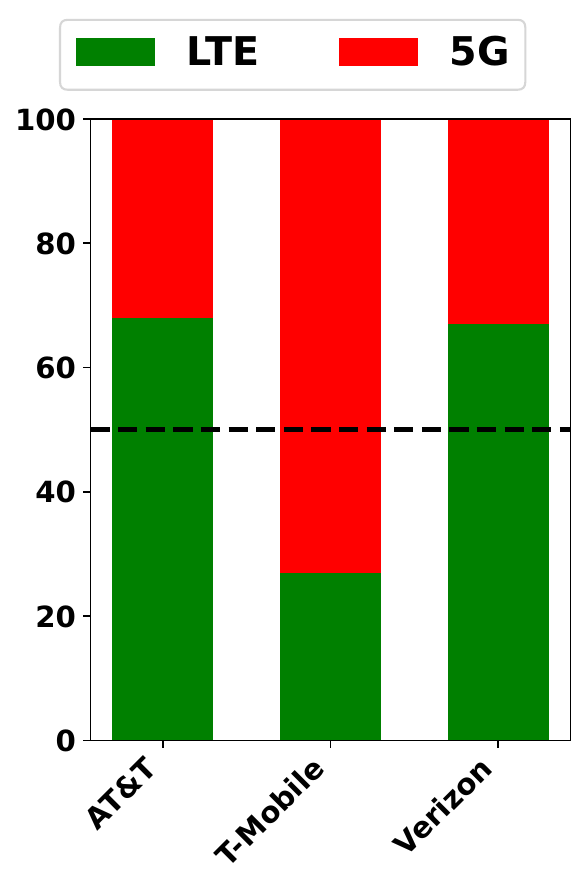}
        \caption{2022.}
        \label{fig:coverage_2022}
    \end{subfigure}  
     \begin{subfigure}[b]{0.32\linewidth}
        \centering
        \includegraphics[width=\linewidth]{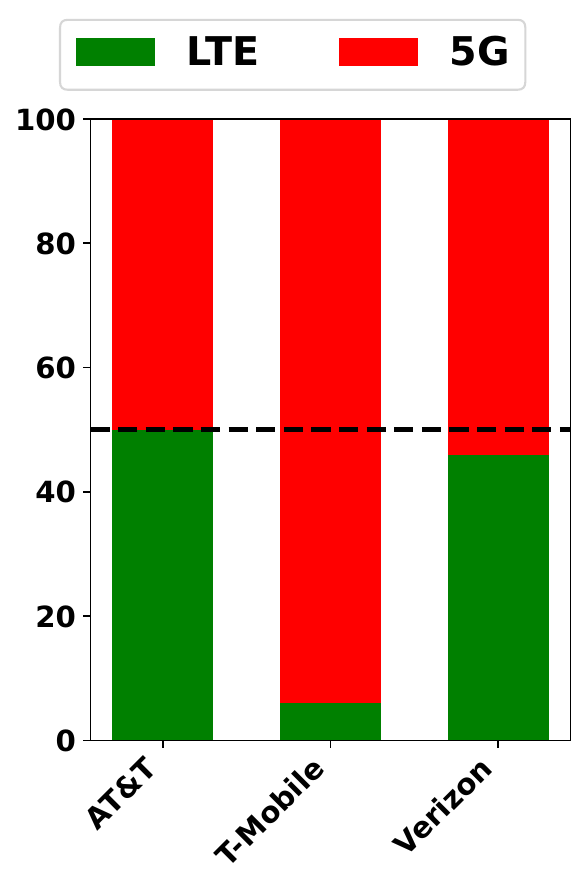}
        \caption{2024.}
        \label{fig:coverage_2024}
    \end{subfigure}  
    \caption{5G and LTE coverage.}
    \label{fig:coverage}
    \vspace{-0.2in}
\end{figure}

\noindent\textbf{Technology coverage.} We define coverage for a particular technology (LTE or 5G) as the fraction of the number of XCAL logs (at a granularity of 1 s) for that technology over the total number of logs. Fig~\ref{fig:coverage_overall} shows that the 5G coverage in our dataset varies from 43.3\% (for Verizon) up to 89.5\% (T-Mobile). Figs.~\ref{fig:coverage_2022}, \ref{fig:coverage_2024} further compare the coverage over the same route in 2022 and 2024, showing an increase in 5G coverage for all 3 operators -- from 32\% to 50\% for AT\&T, from 73\% to 94\% for T-Mobile, and from 33\% to 54\% for Verizon.

\begin{figure}[t]   
     \begin{subfigure}[b]{0.49\columnwidth}
        \centering
        \includegraphics[width=\columnwidth]{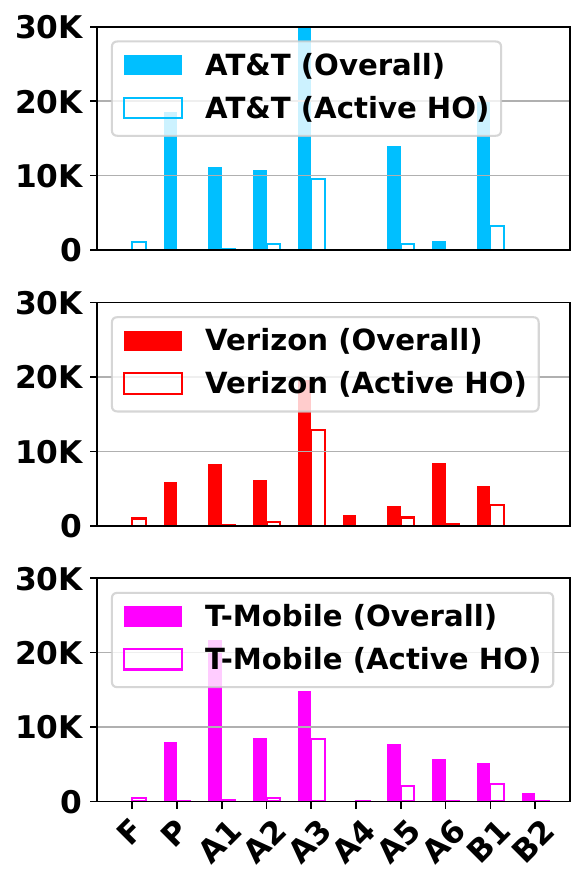}
        \vspace{-0.2in}
        \caption{Total \# of events vs. \# of events that resulted in HOs.}
        \label{fig:overall_comp}
    \end{subfigure}   
     \begin{subfigure}[b]{0.49\columnwidth}
        \centering
        \includegraphics[width=\columnwidth]{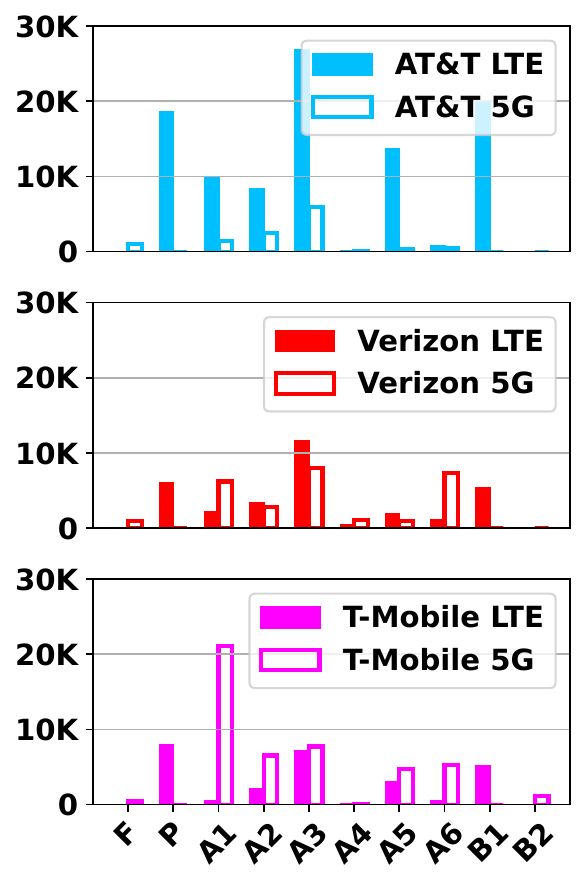}
        \vspace{-0.2in}
        \caption{Event breakdown by technology type.}
        \label{fig:all_events_break_tech}
    \end{subfigure}  
    \vspace{-0.05in}
    \caption{Distribution of HO-related reporting events. F: 5G failures, P: Periodic events.}
    \label{fig:ho_dist}
    \vspace{-0.2in}
\end{figure}

\begin{figure}[t]   
\begin{subfigure}[b]{0.34\columnwidth}
        \centering
        \includegraphics[width=\columnwidth]{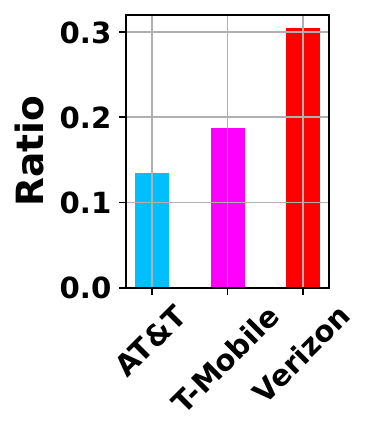}
        \vspace{-0.25in}
        \caption{Overhead.}
        \label{fig:ho_total_events_ratio}
    \end{subfigure} 
     \begin{subfigure}[b]{0.65\columnwidth}
        \centering
        \includegraphics[width=\columnwidth]{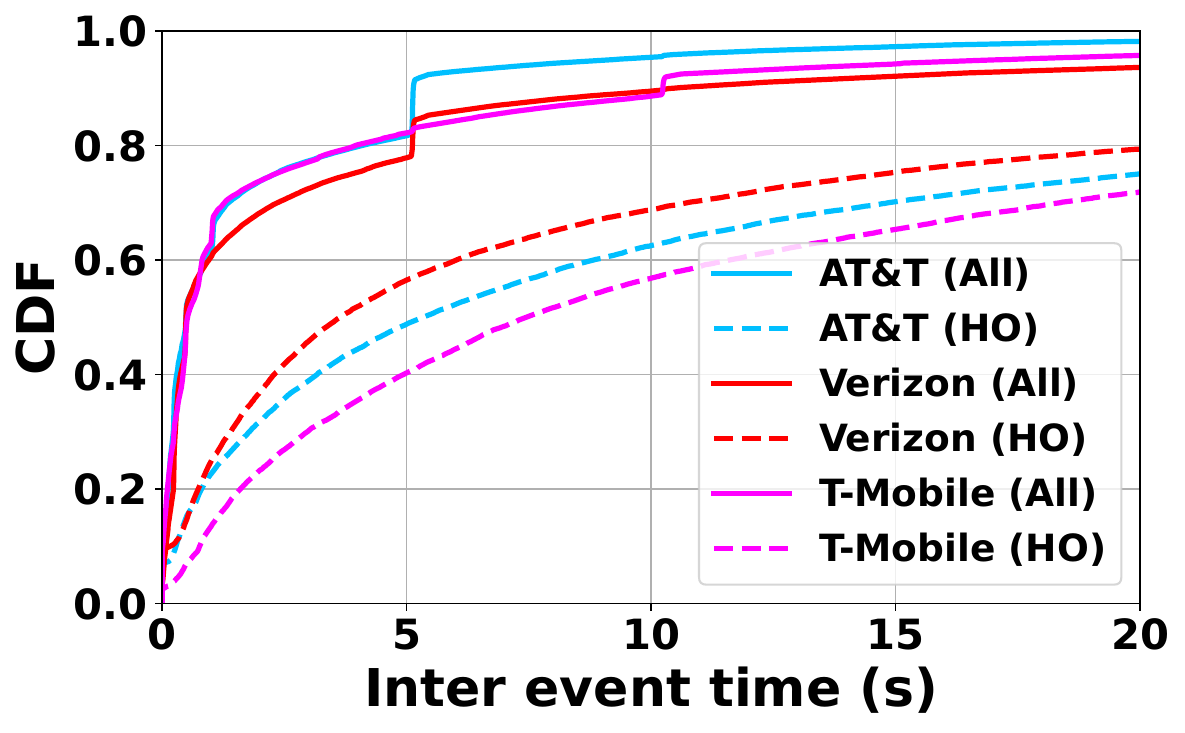}
        \vspace{-0.15in}
        \caption{Inter-MR/HO time.}
        \label{fig:events_inter_mr_cdf}
    \end{subfigure} 
     \vspace{-0.2in}
    \caption{Signaling overhead and inter-MR/HO time.}
    \label{fig:signal_overhead}
    \vspace{-0.25in}
\end{figure}

\noindent\textbf{Distribution of HO-related reporting events.} Fig.~\ref{fig:overall_comp} breaks down the HO-related reporting events for each operator. For each event type, we plot both the total number of events reported by the UE ("Overall") and the number of events that actually resulted in HOs ("Active HO"). 
The three operators generate very different overall numbers of events (108K/58K/73K for AT\&T/Verizon/T-Mobile) and \textit{the distribution of these events is also different for each operator, suggesting drastically different configurations for the event parameters}. A3 is the most popular event for AT\&T and Verizon, while A1 is the most popular event for T-Mobile. 
 AT\&T triggers a much larger number of A5, B1, and periodic events compared to A1 and A2 events. In contrast, the number of A1 events is higher than the number of A5 and B1 events for Verizon and T-Mobile, which also trigger a much smaller number of periodic events compared to AT\&T. 

Fig.~\ref{fig:all_events_break_tech} further breaks down the reporting events according to technology, again showing a large heterogeneity across operators. AT\&T generates a much larger number of LTE events (98K vs. 10K), T-Mobile generates a much larger number of 5G events (26K vs. 47K), while the numbers are more balanced for Verizon (32 K for LTE, 26K for 5G). The numbers appear to be related to the technology coverage (Fig.~\ref{fig:coverage_overall}) for each operator (T-Mobile has much higher 5G coverage than the other two operators), but not necessarily proportionately (e.g, for AT\&T, the coverage ratio is 35\% 5G vs. 65\% LTE, while the event ratio is 9\% 5G vs. 91\% LTE), suggesting again different strategies among operators. For example, T-Mobile triggers many more A1 events in 5G than in LTE, while AT\&T has practically eliminated A5 events in 5G. 
We also observe new events in 5G: B1 (responsible for LTE\textrightarrow 5G HOs), B2 only for T-Mobile (responsible for 5G\textrightarrow LTE HOs in SA), and various 5G failures, but practically no periodic events.

When we look at the events that actually triggered HOs in Fig.~\ref{fig:overall_comp}, we see a very different picture. The total number of events in this case is very similar for the three operators (16K/19K/14K for AT\&T/Verizon/T-Mobile), and we only observe three popular events: A3 (the most popular event for all three operators, accounting for 60-70\% of all the HOs), B1 (15-20\% of all the HOs), and A5 (15\% of all the HOs for T-Mobile, 5-6\% for the other two operators). More importantly, \textit{the number of events that actually triggered HOs is much smaller than the total number of events for all three operators}.

\noindent\textbf{Quantifying the signaling overhead.}
Fig.~\ref{fig:overall_comp} suggests that all three operators use aggressive configurations, forcing the UE to report a very large number of events, but most of the reported events are ignored by the network, resulting in unnecessarily high signaling overhead. Fig.~\ref{fig:ho_total_events_ratio} quantifies this overhead as the ratio of the events that actually triggered HOs compared to the total number of events for each operator; only 13\%/18\%/31\% of the events actually triggered HOs in the case of AT\&T/T-Mobile/Verizon. 

In Fig.~\ref{fig:events_inter_mr_cdf}, we further analyze the signaling overhead by comparing the intervals between successive MRs sent by the UE (inter-MR) and those between successive HOs (inter-HO). The inter-MR interval is under 0.5 s 50\% of the time for all three operators. We also observe distinct “knees” in the CDF -- at 5 s for Verizon and AT\&T, and at 11 s for T-Mobile -- caused by periodic events. In contrast, median inter-HO times range from about 3 s to 7 s, implying a 6–14x higher frequency of MRs compared to actual HOs. The large number of A1 events in Fig.~\ref{fig:overall_comp}, which are used to cancel HOs when the serving cell's signal strength improves, suggest that this overhead could be reduced with more conservative configurations. The absence of periodic events in 5G (Fig.~\ref{fig:all_events_break_tech}) suggests that operators are actually trying to reduce unnecessary signaling on the 5G interface, since periodic events do not typically lead to HOs (see "Active HO" under "P" in Fig.~\ref{fig:overall_comp}).

In summary, \textit{the 5G HO landscape is very different from the LTE landscape; new HO-triggering events have appeared while others have been practically eliminated. The largely different distributions of events across operators suggest drastically different policies in configuring their HO parameters. More importantly, all three operators use aggressive policies on the UE side forcing the UE to report a very large number of events. However, most of these events are finally ignored by the network, resulting in high signaling overhead.}

\section{HO breakdown}
\label{sec:breakdown}

\begin{figure}[t!]
    \centering
    \includegraphics[width=0.7\linewidth]{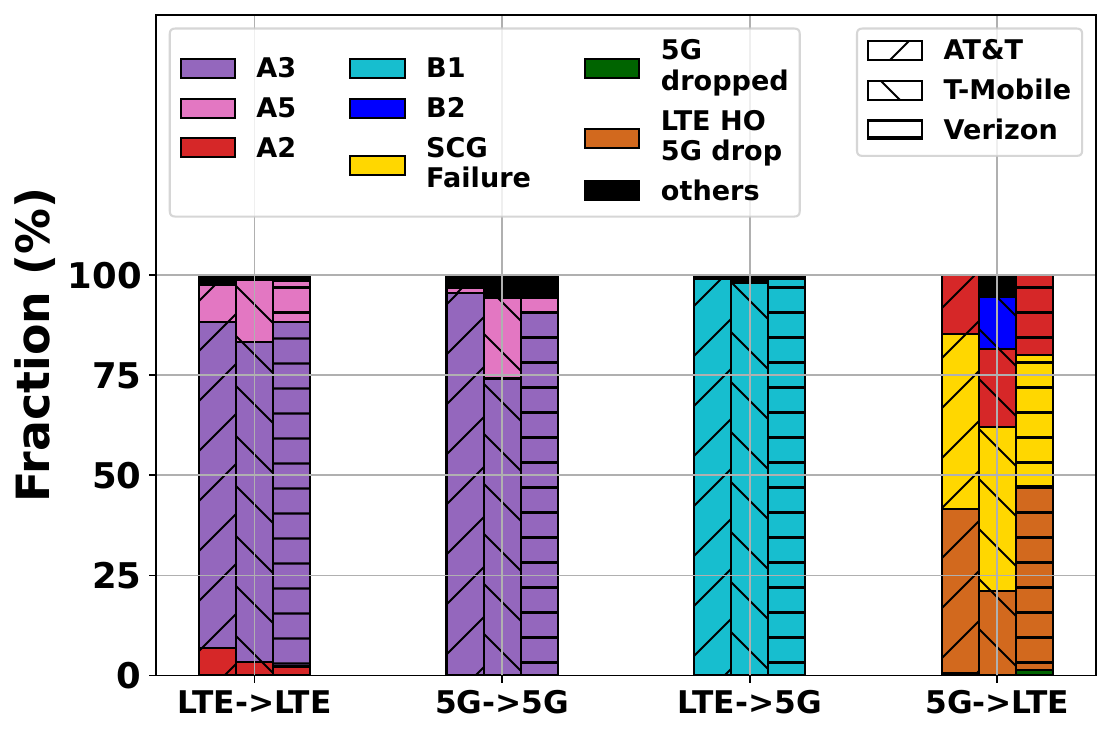}
     \vspace{-0.1in}
     \caption{HO breakdown by reporting events.} 
     \label{fig:breakdown}
      \vspace{-0.2in}
\end{figure}

In this section, we analyze the reporting events responsible for each type of HOs. In addition to LTE\textrightarrow LTE HOs, we have two additional types of HOs in the 5G era: 
5G intra-RAT HOs (e.g., 5G-low\textrightarrow 5G-low, 5G-mid-5G-high) and 5G inter-RAT HOs (LTE\textrightarrow 5G, 5G\textrightarrow LTE). 
Fig.~\ref{fig:breakdown} shows the breakdown of HOs of each HO type
by reporting events.

\begin{figure}[t!]
     \begin{subfigure}[b]{0.48\linewidth}
        \includegraphics[width=\linewidth]{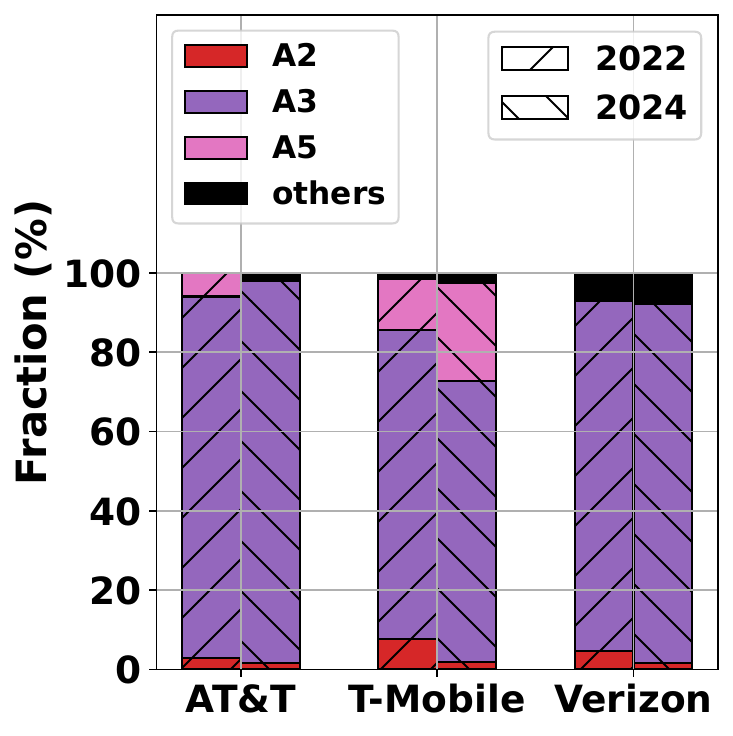}
        \vspace{-0.2in}
        \caption{5G-5G HOs.}
        \label{fig:fiveg_fiveg_2022_2024}
    \end{subfigure}   
     \begin{subfigure}[b]{0.48\linewidth}
        \includegraphics[width=\linewidth]{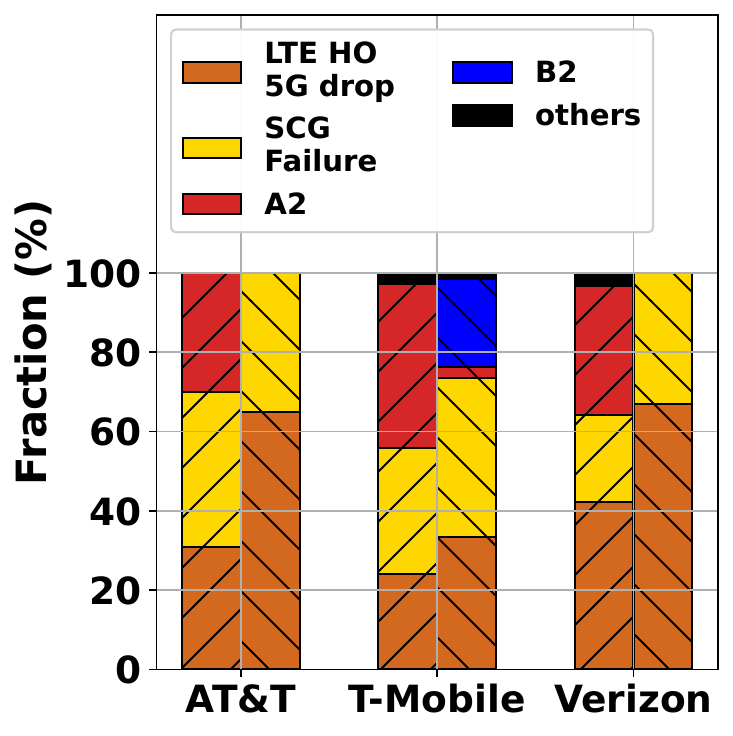}
        \vspace{-0.21in}
        \caption{5G-LTE HOs.}
        \label{fig:fiveg_lte_2022_2024}
    \end{subfigure}    
     \vspace{-0.1in}
     \caption{HO breakdown in 2022 vs. 2024.} 
     \label{fig:breakdown_2022_2024}
      \vspace{-0.3in}
\end{figure}

\noindent\textbf{LTE intra-RAT HOs.} 
The majority of LTE\textrightarrow LTE HOs (80-86\%) for all three operators are triggered by A3 events. Only a small fraction (9-15\%) is triggered by A5, a much smaller fraction (2-7\%) by A2, and an even smaller fraction (3\% or less) by other events (A4, A6, periodic, and even A1 and B1). 

  
\noindent\textbf{5G Intra-RAT HOs.} 
A3 remains the prevalent event for 5G\textrightarrow 5G HOs, being responsible for 74-95\% of all the HOs in all three operators. Interestingly, the percentage of A5 events for AT\&T and Verizon is further reduced to less than 5\%. We found that most intra-band 5G\textrightarrow 5G HOs (67-100\%) are triggered by A3 events. On the other hand, inter-band 5G\textrightarrow 5G HOs are fewer but much more diverse, both across carriers and across bands for the same carrier. Here, A5 is the most popular event, but we also found a non-negligible number of A3, A2, and A4 events.

\noindent\textbf{Inter-RAT HOs.} 
Almost all the LTE\textrightarrow 5G HOs are triggered by B1 events, suggesting that operators are aggressive in upgrading a UE from LTE to 5G as soon as a 5G cell with good signal strength is discovered.\footnote{B1 is triggered when the signal strength of a neighboring cell becomes better than a threshold, regardless of the signal strength of the serving cell.} 
On the other hand, the majority of 5G\textrightarrow LTE HOs are actually due to various types of failures, mostly "SCG failure" and "LTE HO 5G drop". 
The only reporting events here are A2 (15-20\%) and B2 (13\%, only for for T-Mobile SA).

\noindent\textbf{Evolution analysis.} Fig.~\ref{fig:fiveg_fiveg_2022_2024} compares the event distribution for each operator in years 2022 and 2024. We focus only on 5G\textrightarrow 5G and 5G\textrightarrow LTE HOs, since, for the other two HO types, we did not observe any disparity across the two years. 
We observe that T-Mobile exhibits a notable 
rise in A5 usage for 5G\textrightarrow 5G HOs from 2022 to 2024 and B2 events were practically non-existent in 2022 due to the very limited 5G SA deployment, but account for 22\% of the total 5G-LTE HOs for T-Mobile in 2024. More importantly, there are practically no A2-triggered HOs in 2024, while they accounted for 3-8\% of the 5G-5G HOs and 30-41\% of the 5G-LTE HOs in 2022. The lower 5G coverage in 2022 (Fig.~\ref{fig:coverage_2022}) often forced the network to downgrade to LTE due to a lack of neighboring 5G cells or to perform a 5G\textrightarrow 5G HO based on the A2 event itself, if no neighboring 5G cell had sufficiently high signal strength to satisfy the A3 or A5 condition.\footnote{A2 typically does not trigger an HO but simply reports that the serving cell's signal strength has dropped below a threshold, so that the network can start preparing for other events (e.g., A3 or A5).} In contrast, with significantly broader 5G coverage in 2024, an A2 event is typically followed by an A3 or A5r event, resulting in a successful 5G\textrightarrow 5G HO.


\begin{table}[t]
    \captionof{table}{Dominant triggering event per HO type.}
    \label{tab:summary}
    \vspace{-0.1in}
    \resizebox{0.9\columnwidth}{!}{
    \begin{tabular}{|c|c|}
    \hline
    \textbf{HO type}                              & \textbf{Dominant triggering event} \\ \hline
    LTE\textrightarrow LTE       &A3 (80-86\%)                   \\ \hline
    5G\textrightarrow 5G intra-band &A3 (79-96\%)                    \\ \hline
    5G\textrightarrow 5G inter-band 5G  &A5 (52-72\%)               \\ \hline
    LTE\textrightarrow 5G  &B1 (98-99\%)                  \\ \hline
    5G (NSA)\textrightarrow LTE   & 5G failures (78-85\%)                   \\ \hline
    5G (SA)\textrightarrow LTE   & B2 (77\%)                     \\ \hline
    \end{tabular}}
     \vspace{-0.2in}
\end{table}

Table~\ref{tab:summary} summarizes the dominant triggering event for each HO type. \textit{In spite of the increased complexity of the HO landscape in the 5G era due to the existence of many different types of HOs, operators tend to prefer one event for each HO type. However, we also find a large number of other reporting events as well as a number of special cases appearing at a very small percentage. As the 5G HO landscape is still evolving, operators are eliminating certain events (e.g., A2) but relying more on other events (e.g., B2 in 5G SA or A4 in inter-band 5G\textrightarrow 5G HOs). The use of such events may increase in the near future, when 5G deployments will have taken their final form,\footnote{AT\&T and Verizon recently started rolling out 5G SA services.} raising concerns about increased complexity.}



\section{Diversity of HO configurations}
\label{sec:param}

In this section, we conduct an in-depth characterization of the HO configurations extracted from our dataset. 

\noindent\textbf{Diversity metrics.} Similar to~\cite{deng:imc2018}, we characterize the complexity and
diversity of the configuration parameters in terms of three metrics:
(1) Richness ($R$), measured as the number of unique values $m$.
(2) Distribution of the values,
using the Simpson index of diversity $D$,
defined as $D = 1 - \frac{\sum_{i=1}^mn_i(n_i-1)}{N(N-1)}$, where $m$
is the number of unique values, $n_i$ is the count of a single value
$x_i$, and $N$ is the total number of all samples. $D$
takes values in the range $[0,1]$ with lower values indicating less
diversity.
(3) Dispersion over the value range, measured using 
coefficient of variation $C_{\upsilon}$, defined as $C_{\upsilon} =
\frac{\sqrt{Var[X]}}{E[X]}$, where $E[X]$ and $Var[X]$ are the
expectation and variance of the values of parameter X, respectively.
$C_{\upsilon}$ takes values in the range $[0,+\infty]$ with lower values
indicating less dispersion.


\noindent\textbf{Dependence metric.}
To describe the dependence of a parameter $\theta$ on a factor $F$, we use the {\em intersection over union (IoU)} of all the subsets $\{\theta|F = F_j\}$, which shows the fraction of values present in every subset independent of the factor $F_j$. IoU values close to 1 indicate that the parameter $\theta$ is independent of the factor $F$.
In our calculations of IoU, we only consider non-empty subsets. We also ignore parameters that appear in only one subset, as in such cases, IoU defaults to 1, and we mark such parameters as "N/A". If a parameter is marked as "N/A (3)", IoU is not calculated for any operator. 

\begin{figure*}[t!]
\centering
        \includegraphics[width=0.9\textwidth]{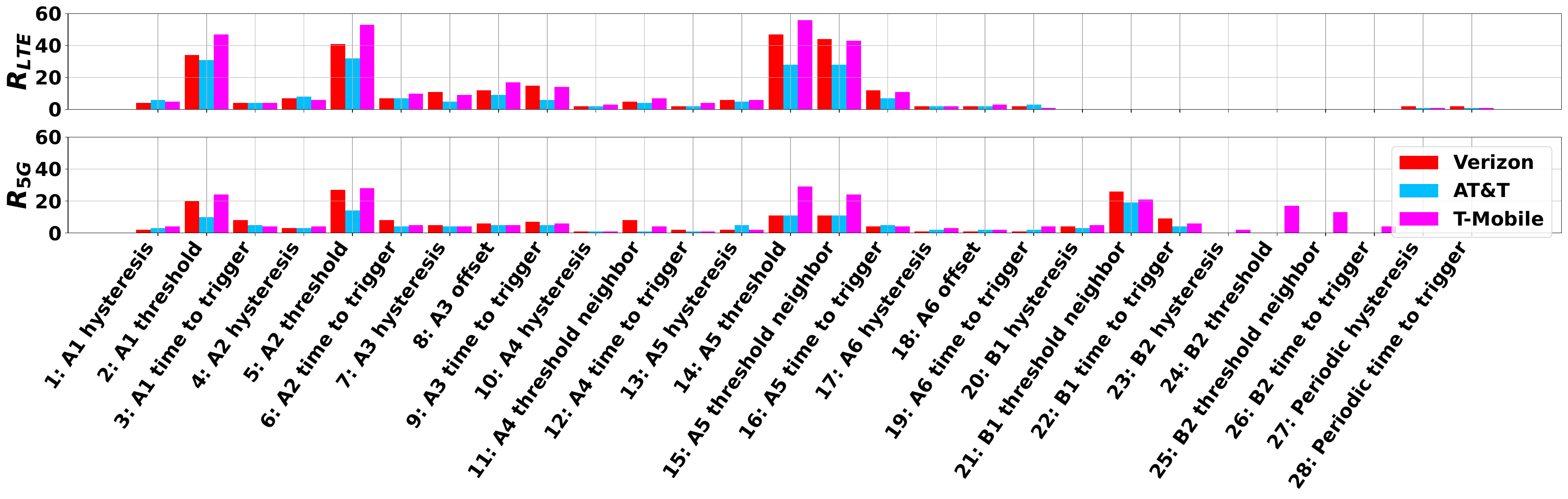}
    \vspace{-0.15in}
    \caption{Richness of HO configuration parameters.}
    \label{fig:richness}
    \vspace{-0.2in}
\end{figure*}

\begin{figure}[t!]
\centering
    \subfloat[Simpson Index.]{
        \includegraphics[width=\columnwidth]{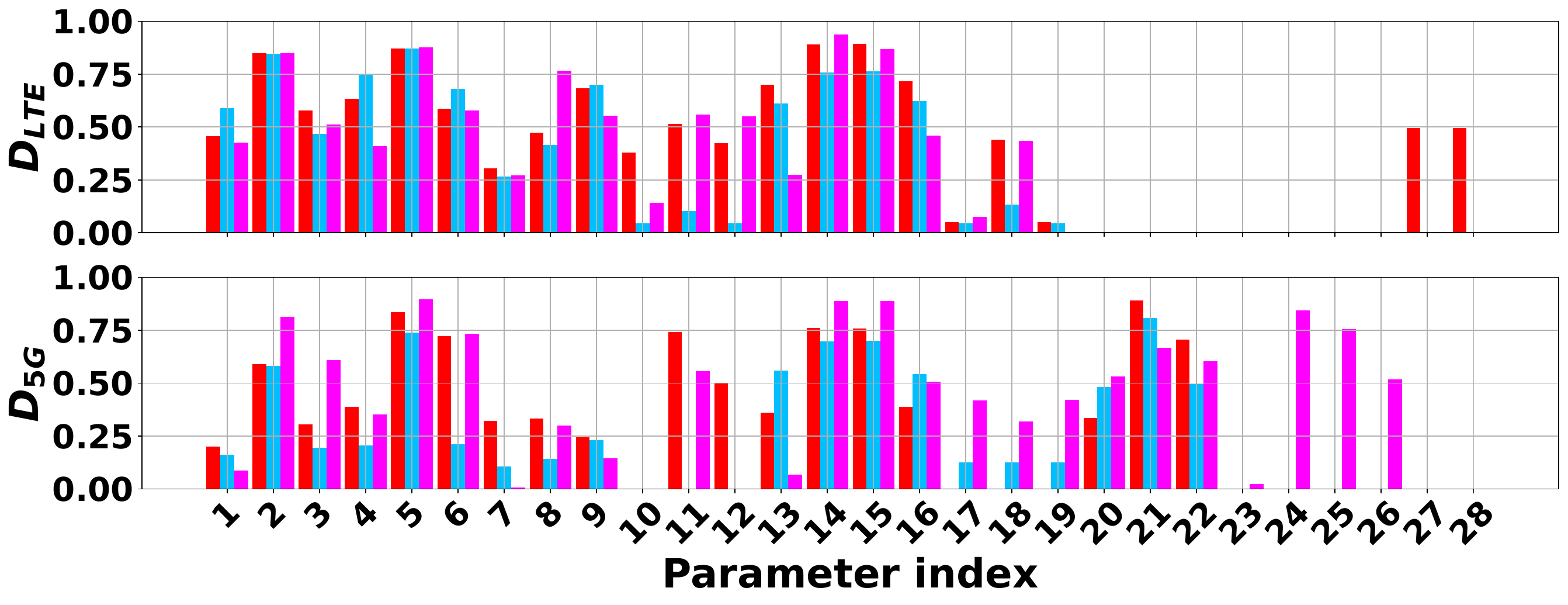}
         \vspace{-0.1in}
        \label{fig:simpson}
   }\\
    \subfloat[Coefficient of variation.]{
        \includegraphics[width=\columnwidth]{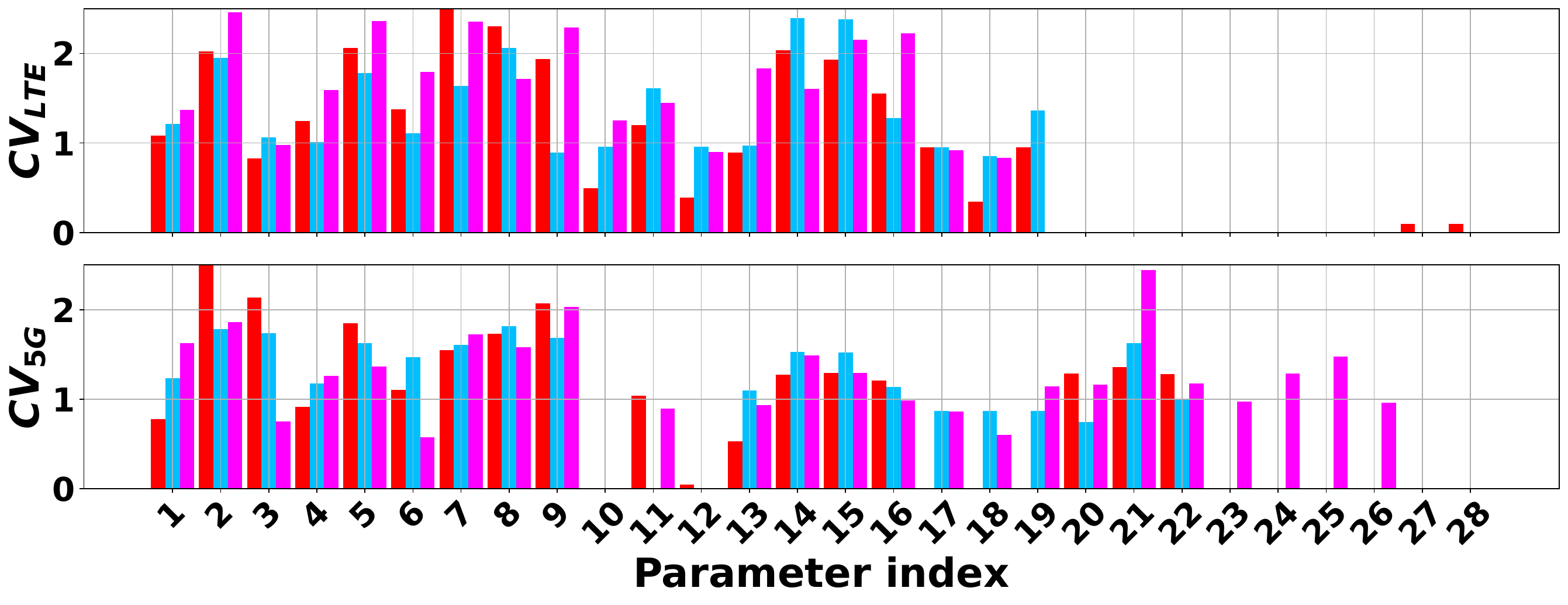}
         \vspace{-0.1in}
        \label{fig:coeff}
    }
    \vspace{-0.1in}
    \caption{Diversity of HO configuration parameters.}
    \vspace{-0.2in}
\end{figure}
\subsection{5G vs. LTE}
\label{sec:param_overall}

Figs.~\ref{fig:richness}, \ref{fig:simpson}, \ref{fig:coeff} show the three diversity metrics -- $R$, $D$, and $C_{\upsilon}$, respectively -- for 28 parameters related to reporting events in LTE and 5G. Note that events B1 and B2 only trigger inter-RAT HOs, and hence, parameters related to those events (20-26) are only shown under 5G.


Fig.~\ref{fig:richness} shows that most LTE parameters take multiple values. On the extreme end, A3 offset has 9-17 distinct values for each operator, and A1 threshold, A2 threshold, and the two A5 thresholds take 28-31 distinct values in AT\&T, 34-44 distinct values in Verizon, and 43-56 distinct values in T-Mobile. These numbers are even higher than those reported for AT\&T in~\cite{deng:imc2018}, where the largest richness observed was 21; even parameters like the A3 hysteresis, which were reported to be single-valued in~\cite{deng:imc2018}, now have 5-11 different values. The only parameters that exhibit low richness are those related to events that are used infrequently (A4, A6, and periodic). Interestingly, T-Mobile exhibits the highest richness values for several parameters even though it has a much lower number of LTE\textrightarrow LTE HOs compared to the other two operators in our dataset suggesting that \textit{richness is not simply an artifact of the number of samples in our dataset but also a result of operator's policies}.

Figs.~\ref{fig:simpson}, \ref{fig:coeff} show that most LTE parameters exhibit a large diversity (much larger than what was reported in~\cite{deng:imc2018}) in terms of both distribution and dispersion. For example, several parameters have $C_\upsilon > 1$ (standard deviation larger than the mean value). This is true even for parameters with low richness, suggesting that, even when operators converge to a small set of values for a parameter, these values have a large range (high $C_\upsilon$) and/or all of them are used at different settings (high $D$), e.g., at different geographic locations.


On the other hand, Fig.~\ref{fig:richness} shows that \textit{the richness of 5G values is much lower, with most parameters having fewer than 12 distinct values}. This is true even for T-Mobile, despite its much higher 5G than LTE coverage (Fig.~\ref{fig:coverage}), reinforcing our conjecture that the number of observed values for a parameter is not related to the number of encountered cells. 

In spite of the low richness, Figs.~\ref{fig:simpson}, \ref{fig:coeff} show that \textit{several 5G parameters exhibit large diversity in terms of distribution} ($D>0.5$) and \textit{almost all of them exhibit large diversity in terms of dispersion} ($C_{\upsilon}$ close to or greater than 1), although still lower compared to LTE. The parameters with large distribution diversity are mostly thresholds and $TTT$s, while most hysteresis parameters have low diversity, although there are some exceptions. 
In contrast to LTE, several 5G parameters with low richness have high distribution diversity; several of them have different values in different geographic locations and/or different 5G bands, in particular mmWave. We also observe parameters with large distribution diversity  but low dispersion diversity and vice versa. 

\begin{figure}[t]
    \centering
    \includegraphics[width=1\columnwidth]{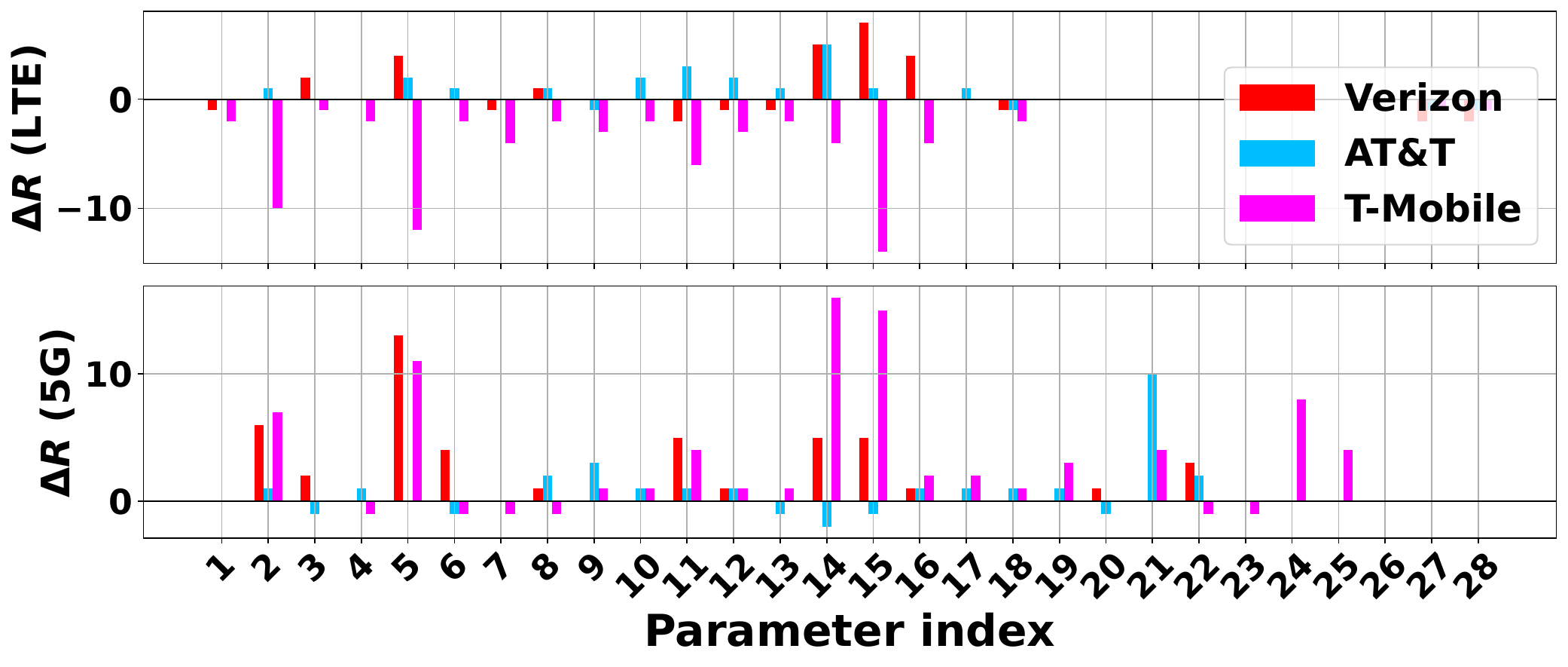}
    \vspace{-0.25in}
    \caption{Richness comparison, R(2024) - R(2022).}
    \label{fig:richness_2022_2024}
    \vspace{-0.25in}
\end{figure}

\noindent\textbf{Evolution analysis.}
Fig.~\ref{fig:richness_2022_2024} shows the difference in richness for each parameter in 2022 vs. 2024, $\Delta R = R(2024) - R(2022)$. $\Delta R > 0$ for a given parameter indicates that the operator uses more values in 2024 than in 2022; $\Delta R < 0$ indicates the opposite. We observe distinct trends for LTE and 5G.

In the case of LTE, $\Delta R$ is 0 or negative for most parameters. $\Delta R$=0 suggests that operators have mostly stopped tweaking the LTE configurations as LTE networks are matured. A negative $\Delta R$ on the other hand is probably due to the fact that we encountered a significantly smaller number of LTE cells in 2024 compared to 2022, i.e., coverage here starts playing a role in diversity. This is true in particular for T-Mobile, which has very low LTE coverage in 2024 (Fig.~\ref{fig:coverage_2024}) and exhibits the largest reduction in richness in Fig.~\ref{fig:richness_2022_2024}. 

In contrast, in the case of 5G, we observe mostly positive $\Delta R$ suggesting that operators have diversified their configurations significantly during the past 2 years (in some cases, by adding more than 10 new values), to adapt to the evolving traffic demands and the extended coverage. The parameters with the largest positive $\Delta R$ are often different for different operators, suggesting that operators prioritize different aspects of 5G HOs. 
Interestingly, the parameters of the most popular event A3 (indexes 7, 8, 9) show only minor variations in their values over the 3-year period (in fact, Verizon did not modify two of these parameters and T-Mobile reduced the richness for two of them).


\begin{figure}[t]
    \begin{subfigure}[b]{0.9\columnwidth}
        \includegraphics[width=\columnwidth]{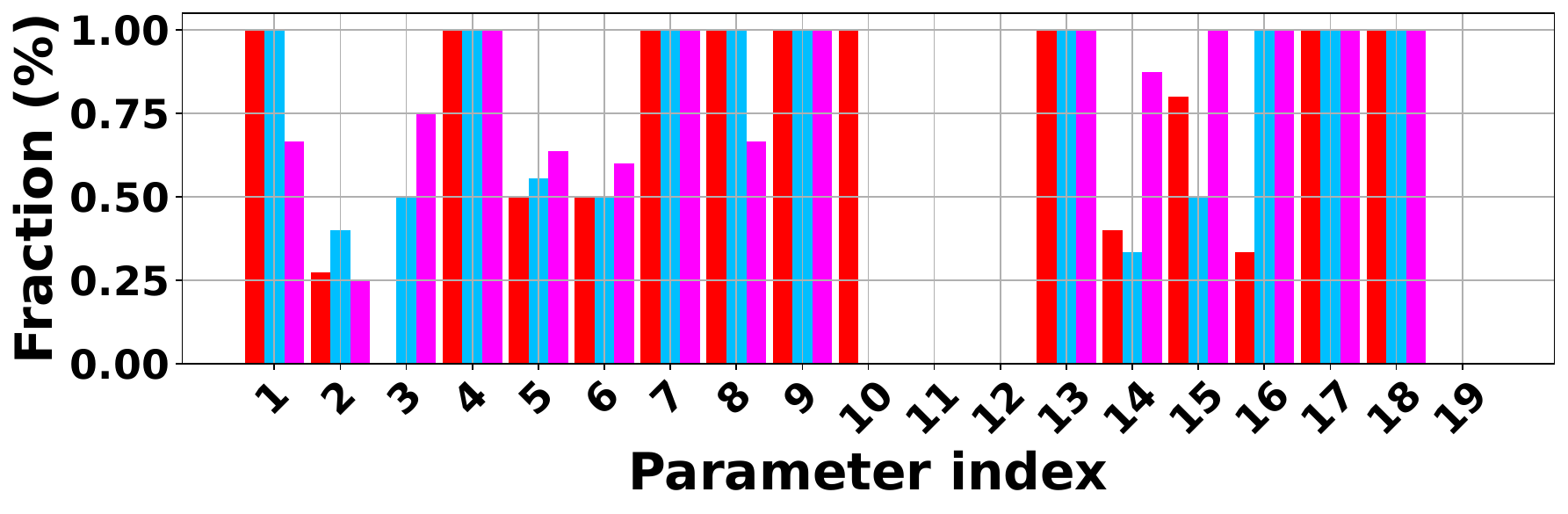}
        \vspace{-0.2in}
        \caption{2022.}
        \label{fig:5gfrac_2022}
    \end{subfigure}
    \begin{subfigure}[b]{0.9\columnwidth}
        \includegraphics[width=\columnwidth]{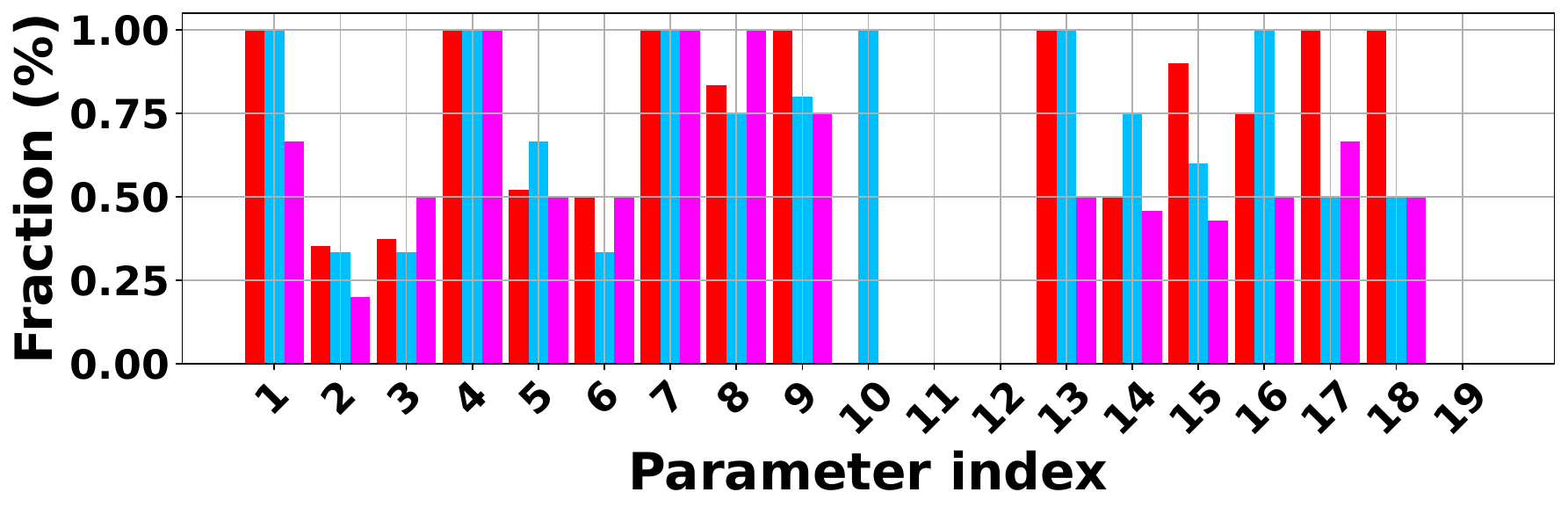}
        \vspace{-0.2in}
        \caption{2024.}
        \label{fig:5gfrac_2024}
    \end{subfigure}
    \vspace{-0.1in}
    \caption{Fraction of 5G values for each parameter that are also used in their LTE counterpart.}
    \label{fig:5gfrac}
    \vspace{-0.1in}
\end{figure}

Given the comparatively smaller richness of 5G parameters compared to their LTE counterparts in Fig.~\ref{fig:richness}, a relevant question is whether operators use a subset of the LTE values for their 5G configurations or a completely different set of values. To answer this question, we plot in Fig.~\ref{fig:5gfrac} the fraction of 5G values for each parameter that are also used in the same LTE parameter for 2022 and 2024. In 2022, we observe a moderate to high overlap of 5G and LTE parameters with percentage $\geq$ 50\% for most parameters and often equal to 100\%. 
In 2024, the overlap has reduced for several parameters in the case of T-Mobile (11) and AT\&T (7), but increased or remained the same for most parameters (14) in the case of Verizon.
While all three operators initially relied on the experience gained from their LTE configurations to configure their 5G deployments, T-Mobile and AT\&T have started diverging from their LTE configurations, while Verizon keeps adding new values from the pool of LTE values.


In summary, \textit{operators use very diverse LTE and 5G configurations with several parameters exhibiting high degrees of diversity in terms of richness, distribution, and dispersion of their values. HO parameters are configured differently by each operator, with most of them exhibiting low to moderate overlap in their values across operators. Operators have stopped tweaking their LTE configurations, but continue fine-tuning their 5G configurations over the past 2 years and increasing their richness, although they prioritize different types of HOs and adopt different strategies in the selection of parameter values.}

\begin{figure}[t]
\centering
    \subfloat[IoU.]{
        \includegraphics[width=0.48\textwidth]{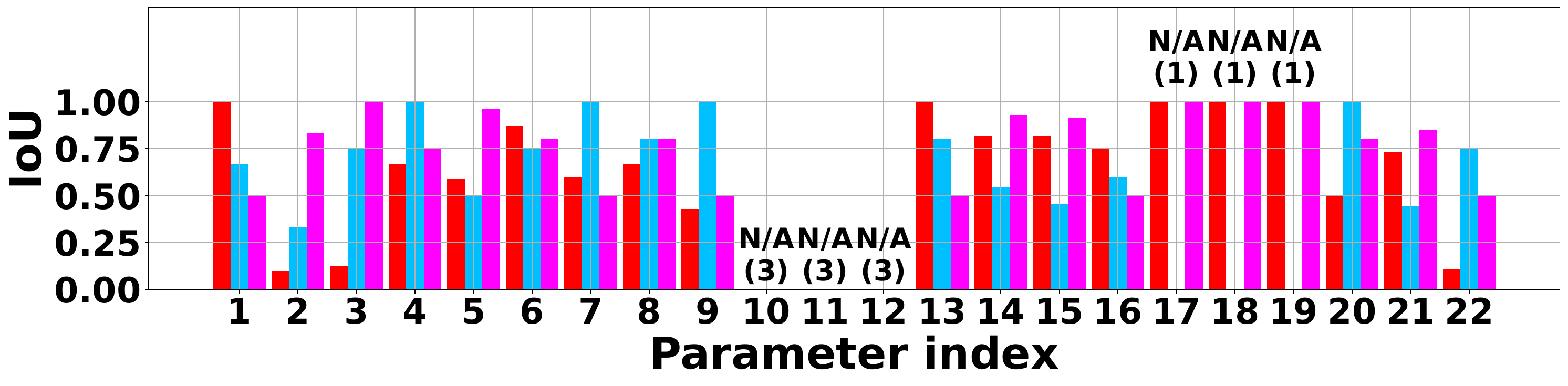}
         \vspace{-0.1in}
        \label{iouband}
    }\\
    \subfloat[IoU(2024) - IoU(2022).]{
        \includegraphics[width=0.48\textwidth]{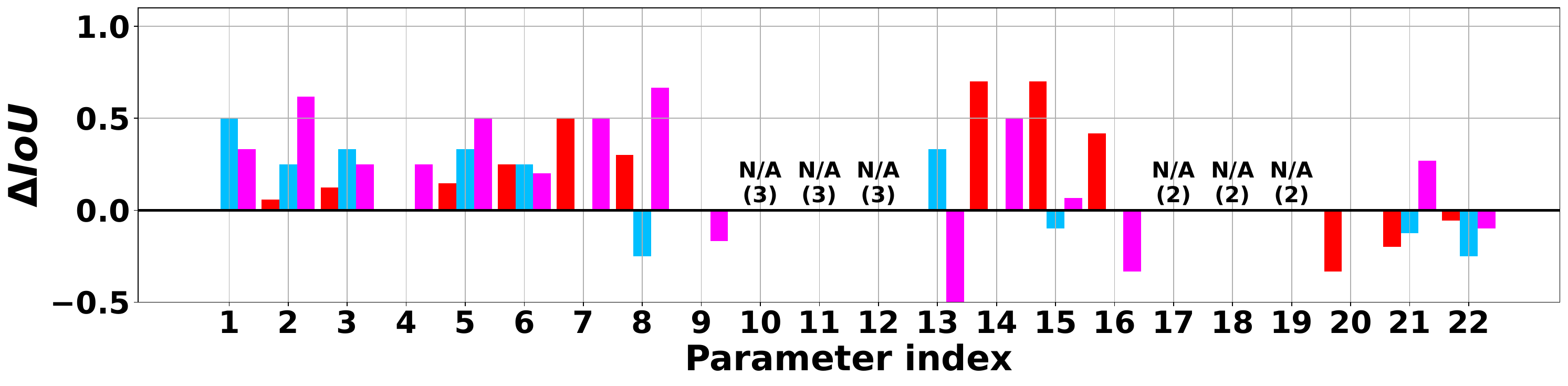}
         \vspace{-0.1in}
        \label{fig:iou_5g_2022_2024}
    }
\vspace{-0.1in}
    \caption{Dependence on 5G band.}
    \vspace{-0.1in}
\end{figure}

\subsection{Dependence on 5G band and architecture}
\label{sec:param_band}

We explore the dependence of the 5G configurations on the 5G
frequency band (low, mid, or high) in Fig.~\ref{iouband}. For AT\&T and T-Mobile, we only
consider the low and mid bands, since we have very few instances
of 5G mmWave HOs.

Fig.~\ref{iouband} shows that Verizon has the largest number of parameters with $IoU > 0.5$ (14/19 vs. 11 for AT\&T and 13 for T-Mobile. This is rather surprising, taking into account the fact that Verizon's IoU is calculated over 3 bands. 
At the same time, Verizon has the 3 parameters with the lowest IoU, suggesting that the operator uses largely different criteria in each band to determine when to cancel an HO (A1, indexes 2 and 3) or to promote a UE to 5G (B1, index 22). 
For all three operators, the parameters with the largest IoU values are either hysteresis and TTT values or parameters of events used infrequently (A6, A2). Interestingly for AT\&T and T-Mobile, the A3 offset (index 8) also has very high IoU.


\begin{figure}[t]
    \centering
    \includegraphics[width=1\columnwidth]{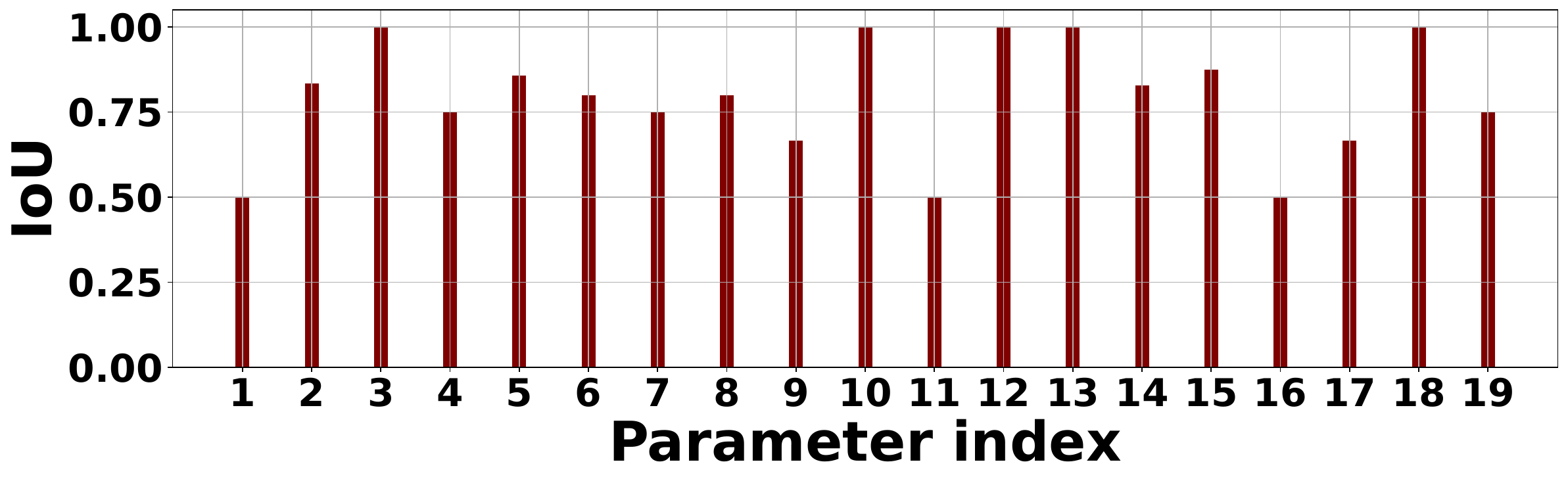}
    \vspace{-0.2in}
    \caption{IoU comparison for 5G SA vs. NSA.}
    \label{fig:sa_nsa}
    \vspace{-0.1in}
\end{figure}

We also explore the dependence of HO configurations on the 5G architecture (NSA vs. SA) in Fig.~\ref{fig:sa_nsa} using only the T-Mobile dataset, as this is the only operator that supports SA.
We observe that all parameters have $IoU > 0.5$ and 14 out of 19 parameters have $IoU \geq 0.75$, indicating in general \textit{independence of the 5G architecture}. 

\noindent\textbf{Evolution analysis.} Fig.~\ref{fig:iou_5g_2022_2024} plots the difference in IoU, $\Delta IoU$, between 2024 and 2022 for each parameter. We observe that most parameters have positive $\Delta IoU$, indicating that, as the technology matures, differences across bands tend to be eliminated with operators converging to a subset of values for each parameter that is used across all bands. However, there are a few exceptions, different for each operator, suggesting again different strategies for different operators in fine-tuning certain parameters. For example, the parameters with the lowest (negative) $\Delta IoU$ are 13 and 16 for T-Mobile and 20, 21 for Verizon, indicating a focus of the operator on fine-tuning A5 and B1 events, respectively. 
We omit the temporal analysis for the dependence on the 5G architecture, since in 2022 we had only 29 5G SA HOs. (2\% of the total).

Overall, \textit{as the 5G technology matures, operators are converging to a set of values for each parameter that are used across all 5G bands. Hysteresis and TTT parameters, as well as parameters of events triggered infrequently, typically have the largest overlap in their values across bands.} 
\subsection{Dependence on area type}
\label{sec:param_geo}

\begin{figure}[t]
    \subfloat[IoU.]{
        \includegraphics[width=0.48\textwidth]{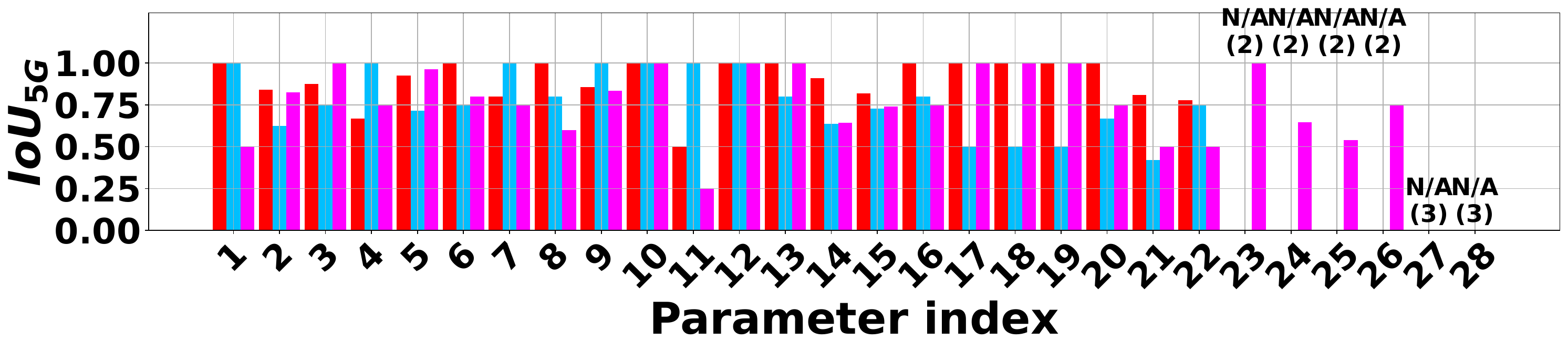}
        \label{fig:ioucity}
    }\\
    \subfloat[IoU(2024) - IoU(2022).]{
        \includegraphics[width=0.48\textwidth]{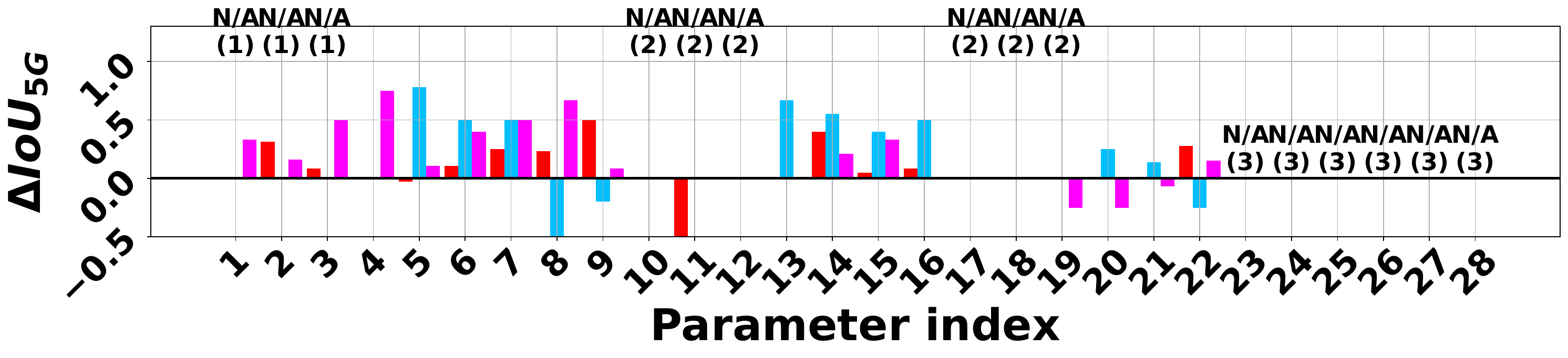}
        \label{fig:ioucity_2022_2024}
    }
    \vspace{-0.1in}
    \caption{Dependence of 5G parameters on area type (large city vs. otherwise).}
    \label{fig:iou_city_tz_5g}
    \vspace{-0.1in}
\end{figure}

\begin{figure}[t]
    \centering
    \includegraphics[width=0.5\textwidth]{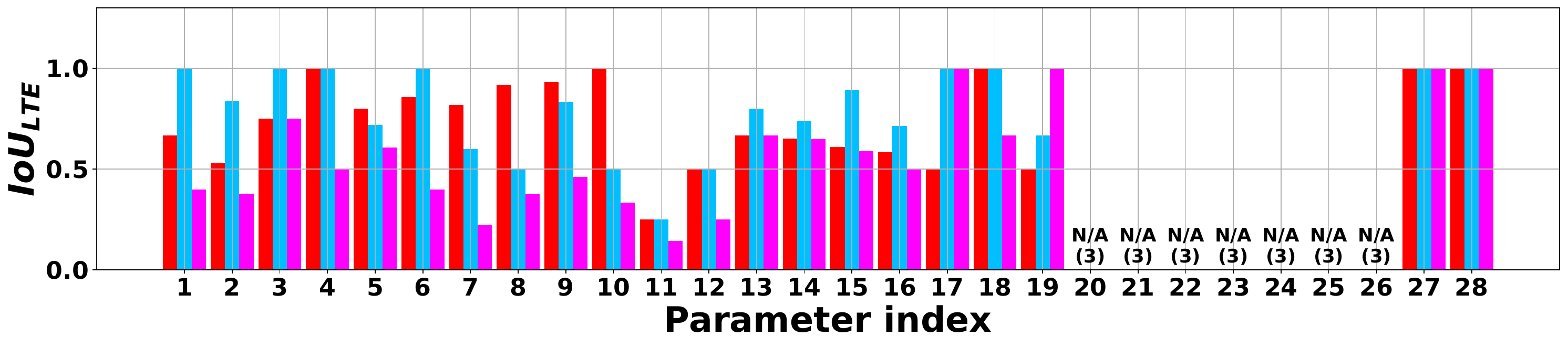}
    \vspace{-0.1in}
    \caption{Dependence of LTE parameters on area type (large city vs. otherwise).}
    \label{fig:ioucity_app}
    \vspace{-0.2in}
\end{figure}

We explore whether operators use different policies in large urban areas, which typically have a much denser BS deployment, compared to the rest of the country. We identify all cities with a population greater than 300K in our dataset and we compute the IoU for each parameter (Fig.~\ref{fig:ioucity}) between regions within a 3-mile radius and outside a 10-mile radius from the downtown of any large city.\footnote{Since the base station locations are not available, we use the UE locations. To avoid misclassifications, when the UE is outside the 3-mile radius of a large city, but is connected to a base station inside the 3-mile radius, we omit all the locations between a 3-mile and 10-mile radius.} 

Surprisingly, Fig.~\ref{fig:ioucity} shows a very large
overlap in the values of many 5G parameters. 21/18/22 parameters have $IoU > 0.5$ and 11/7/8 parameters have $IoU = 1$ for Verizon/AT\&T,/T-Mobile, respectively. The parameters with different values depending on the area type are mainly from events that rarely trigger HOs (A1, A4, A6), with the exception of B1 parameters (for AT\&T and T-Mobile). In contrast, the overlap is much smaller in the case of LTE (Fig.~\ref{fig:ioucity_app}) for all three operators, with only 16/16/10 parameters having $IoU > 0.5$ and 5/6/4 parameters having $IoU = 1$.

Given that the LTE technology is much more mature, one may conjecture that, in the coming years, we will start observing a similar differentiation for 5G, as operators keep fine-tuning their configurations. However, comparing the IoU values in 2022 vs. 2024 in Fig.~\ref{fig:ioucity_2022_2024} shows the opposite trend. Most parameters have a positive $\Delta IoU$ indicating that operators tend to converge to the same set of values regardless of the area type. 

Overall, \textit{operators often select different configurations inside and outside large urban centers for their LTE parameters, but use largely overlapping configurations for their 5G parameters. 
}

\section{Impact on Performance}
\label{sec:performance}

In this section, we study the impact of HO configurations on performance. While previous works~\cite{hassan:sigcomm2022,xu:sigcomm:2020,ghoshal:imc2023} have shown that 5G HOs can have a detrimental impact on throughput during their execution stage and several types of 5G HOs do not always improve signal strength and throughput, our goal here is to analyze and understand the {\em impact of specific 5G HO configurations} employed by operators on performance. 

In the case of intra-RAT (5G to 5G) HOs, we focus on the two main events: A3 and A5. In the case of inter-RAT HOs, we focus on B1 events used for LTE\textrightarrow 5G HOs. Recall from Fig.~\ref{fig:breakdown} that 5G\textrightarrow LTE HOs occur mostly due to various types of 5G failures. Among the two events that trigger 5G\textrightarrow LTE HOs, A2 is no longer used in 2024 (Fig.~\ref{fig:fiveg_lte_2022_2024}), while B2 is only used by T-Mobile in 5G SA, and we only have a total of 88 B2-triggered HOs in our dataset -- too few for a meaningful analysis.


As we saw in \S\ref{sec:param}, many parameters have a very high richness, resulting in a very large number of possible configurations for a given event. For example, in the case of A3 for T-Mobile 5G, there is a total of 4x5x6 = 120 possible combinations of hysteresis, offset, and $TTT$ (Fig.~\ref{fig:richness}). In the following, for each event, we focus on the \textit{dominant} configurations, which together are responsible for at least 90\% of the HOs triggered by that event.

\subsection{Impact on signal strength and ping pong HOs}
\label{sec:rsrp}
  
We select signal strength (RSRP) and the fraction of wasteful HOs (also known as "ping pong HOs") as the primary metrics of interest, as
these are the metrics that operators can
control by selecting appropriate values for various configuration
parameters (thresholds, offsets, $TTT$, etc.), and they allow
us to directly quantify the impact of selected configurations on performance. 

For a pair of consecutive HOs, $HO_n$ from $PCI_{n-1}$ to $PCI_n$ and $HO_{n+1}$ from $PCI_n$ to $PCI_{n+1}$, occurring within a short period of time (15 s in our analysis), $HO_{n+1}$ is a ping pong HO, if $PCI_{n+1} = PCI_{n-1}$.
In the case of RSRP, we are interested
in both the RSRP difference right after and right before a HO ($\Delta
RSRP = RSRP_{post}-RSRP_{pre}$) and the absolute RSRP right before a HO ($RSRP_{pre}$). The
former quantifies the improvement in user performance after a HO,
while the latter quantifies the degradation of user performance before
a HO. 

The choice of values for the main HO configuration parameters introduces a \textit{tradeoff} to the operators. For example, in the case of A3 configurations, a larger $\Delta_{A_3} + H_{A_3}$ (Table~\ref{tab:meas_events}) can result in better post-HO performance at the cost of poorer performance before the HO. Similarly, a longer $TTT$ increases confidence in the reported measurements and can lead to better post-HO performance, while reducing the probability for a ping-pong HO, but it can delay the HO decision resulting in poorer pre-HO performance.

\begin{figure}[t]
\centering
    \includegraphics[width=\columnwidth]{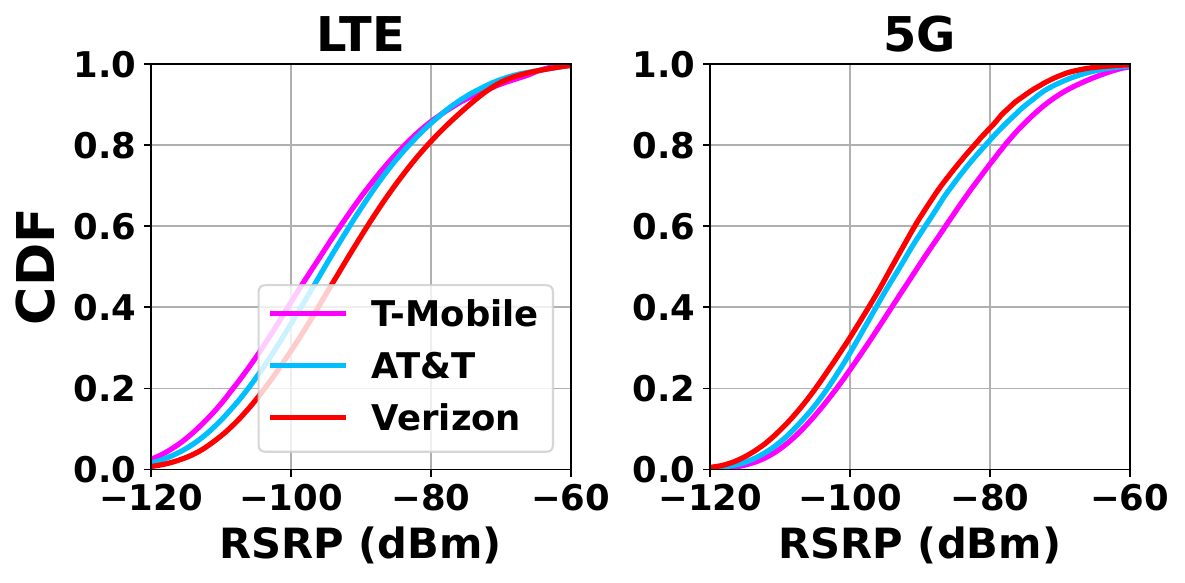}
     \vspace{-0.2in}
    \caption{CDFs of LTE and 5G RSRP for the three operators in our dataset.}
    \label{fig:rsrp_dist}
    \vspace{-0.1in}
 \end{figure}
 
However, there is a subtlety in using $RSRP_{pre}$ to compare performance across operators. Unlike $\Delta RSRP$, which is a relative metric, $RSRP_{pre}$ is an absolute metric and its value depends not only on the HO decisions, but also on the overall cellular coverage. In other words, it is possible that an operator X using a longer $TTT$ may still achieve a higher $RSRP_{pre}$ than an operator Y using a shorter $TTT$, e.g., if operator X has a denser deployment and hence better coverage or uses higher Tx power than operator Y. Indeed, we found (see Fig.~\ref{fig:rsrp_dist}) that the three operators have different RSRP distributions, as a result of different deployment strategies~\cite{ghoshal:imc2023}.



\subsubsection{A3 HOs}
\label{sec:a3hos}

\begin{figure}[t]
	\centering
        \subfloat[Dominant combinations $(\Delta_{A_3} + H_{A_3}, TTT)$.] {
\label{tab:a3_intra}
\resizebox{0.35\textwidth}{!}{
\begin{tabular}{|c|c|c|}
\hline
\textbf{AT\&T}   & \textbf{T-Mobile}  &\textbf{Verizon}  \\ \hline
 \begin{tabular}[c]{@{}c@{}}(8, 640): 89\%\\ (4, 320): 7\%\\ Count: 2\end{tabular} & \begin{tabular}[c]{@{}c@{}}(8, 640): 53\%\\ (10, 640): 33\%\\ (10, 320): 10\%\\ Count: 3\end{tabular} & \begin{tabular}[c]{@{}c@{}}(6, 256): 72\%\\ (8, 256): 14\%\\ (8, 480): 7\%\\ Count:3\end{tabular} \\ \hline
\end{tabular}
}} \\
	\subfloat[Performance.] {
        \includegraphics[width=0.4\textwidth]{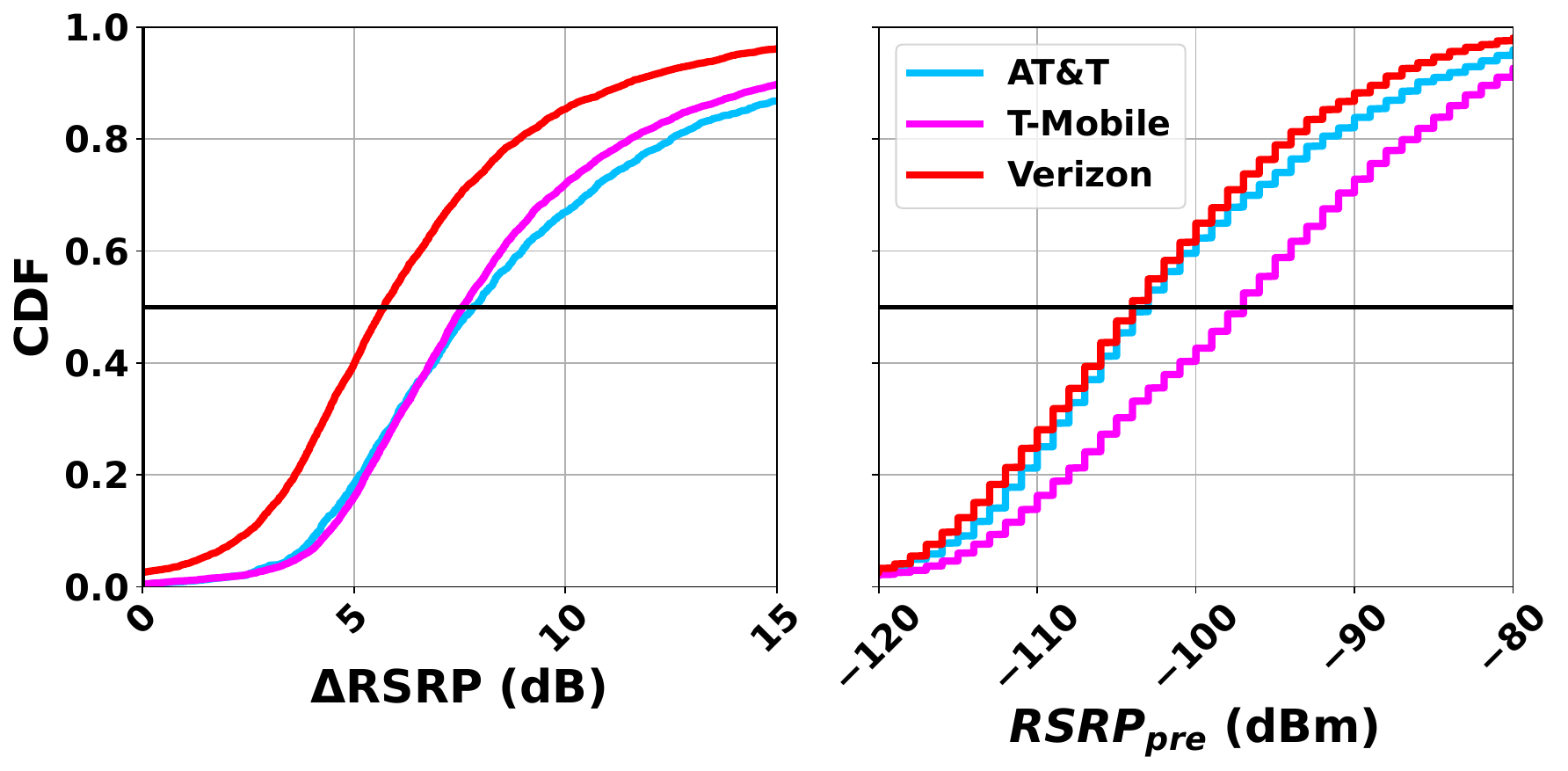}
        \label{fig:perf_intra_a3}
	}
	\vspace{-0.15in}
	\caption{Dominant parameter combinations and RSRP analysis for the A3 event.}
	\label{fig:a3_tab_rsrp}
    \vspace{-0.2in}
\end{figure}

An A3 event is triggered when the RSRP of a neighboring
cell remains $\Delta_{A_3} + H_{A_3}$ dB
higher than the RSRP of the serving cell for $TTT$ ms. Fig.~\ref{tab:a3_intra} shows the dominant configurations --
tuples of $(\Delta_{A_3} + H_{A_3}, TTT)$ -- for all three operators. AT\&T has 2 dominant configurations, responsible for 96\% of the HOs, while T-Mobile and Verizon each have 3 dominant configurations, responsible for 96\% and 93\% of the HOs, respectively. The configurations are different across operators.

\begin{figure}[t]
    \subfloat[$\Delta_{RSRP}$ parameter breakdown.]{
        \includegraphics[width=0.48\textwidth]{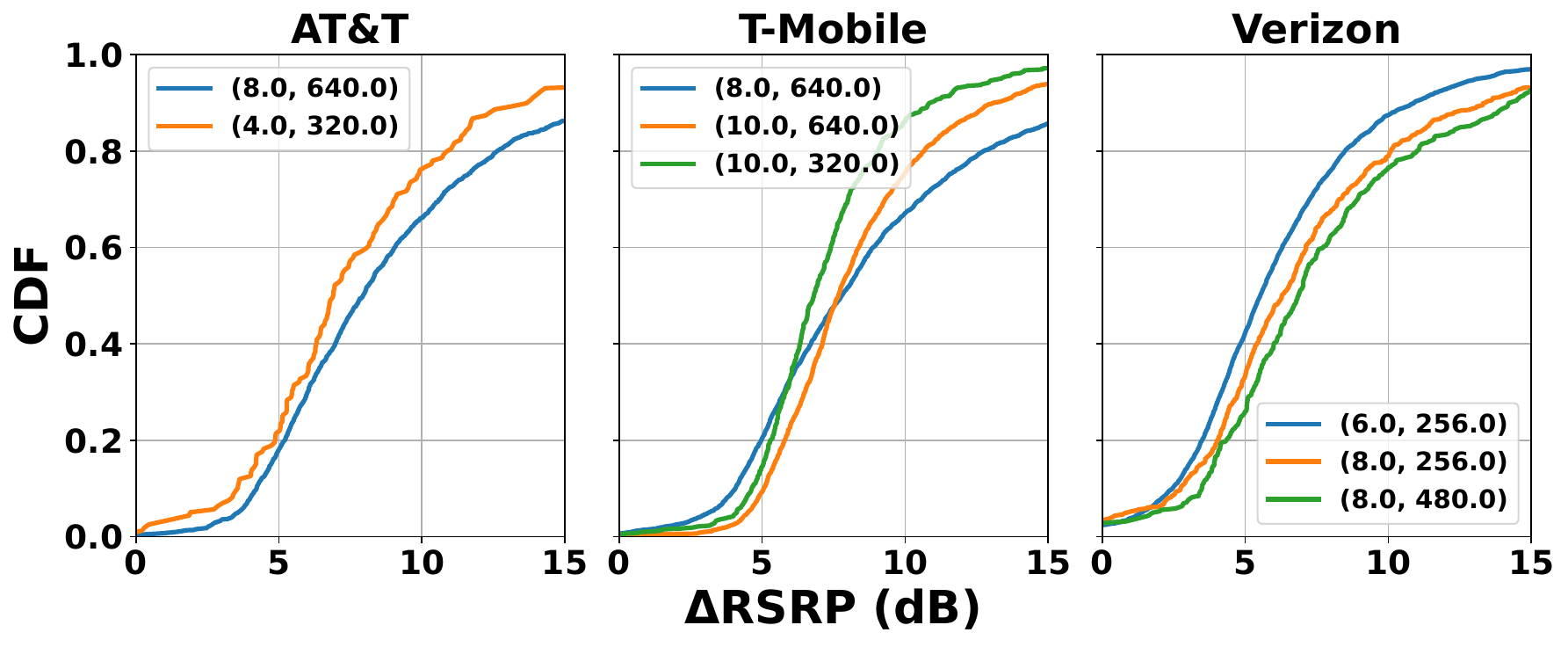}
        \label{fig:a3_param_rsrp_diff}
    }\\
\subfloat[$RSRP_{pre}$ (dBm) parameter breakdown.]{
        \includegraphics[width=0.48\textwidth]{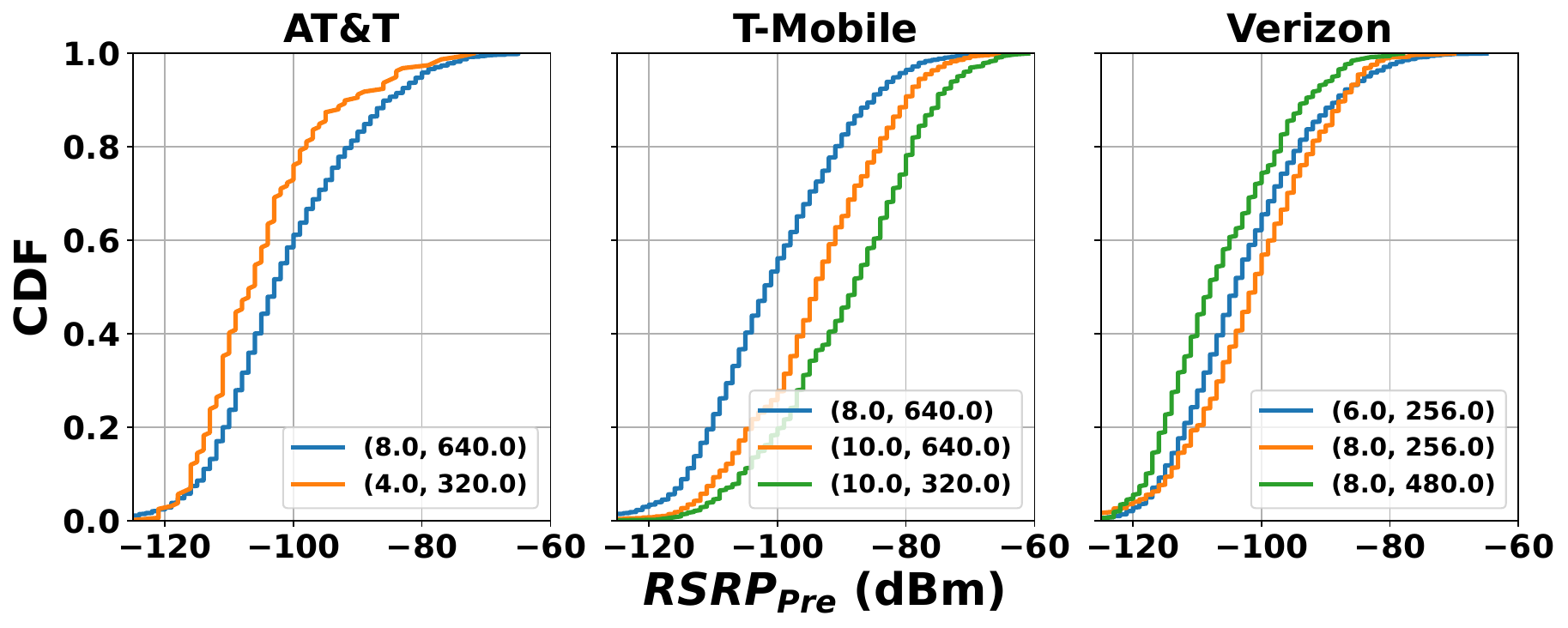}
        \label{fig:a3_param_rsrp_before}
    }\\
    \subfloat[$RSRP_{neighbor}$ parameter breakdown.]{
        \includegraphics[width=0.48\textwidth]{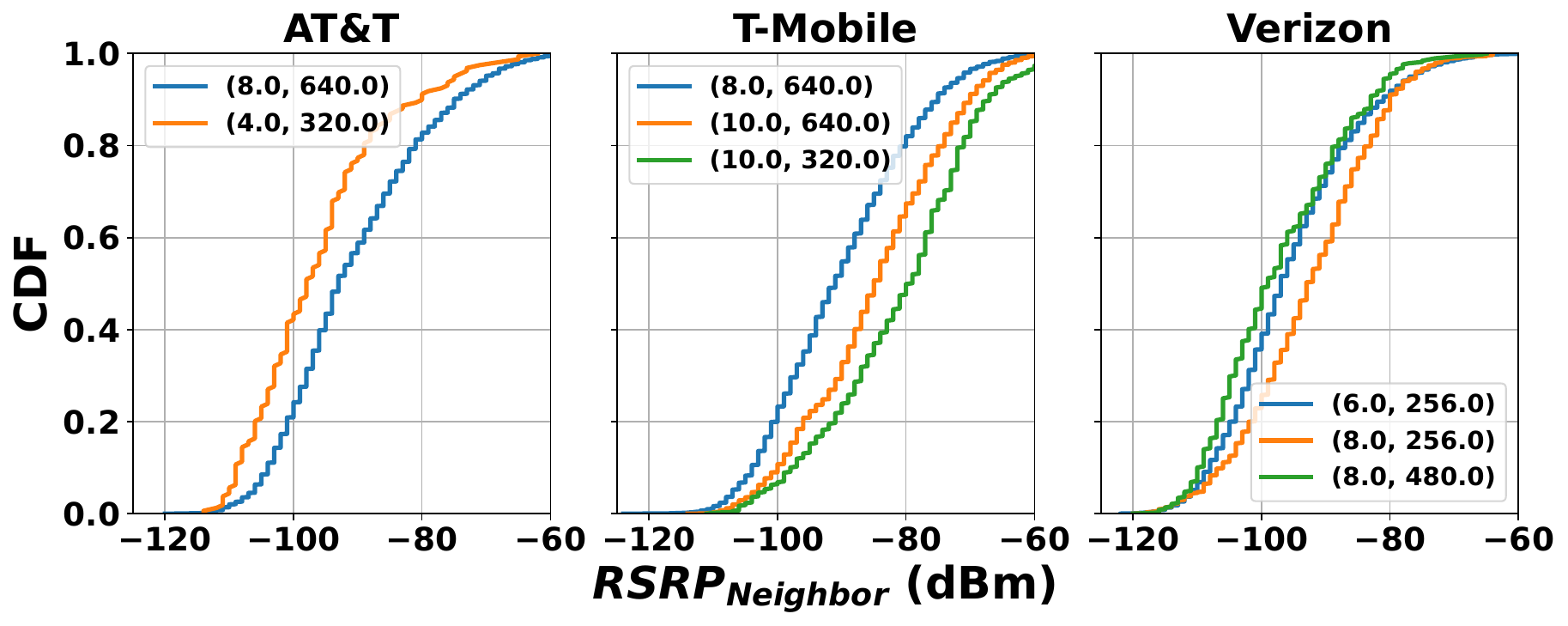}
        \label{fig:a3_param_rsrp_neighbor}
    }
    \vspace{-0.1in}
    \caption{Impact of A3 configurations on performance.}
     \label{fig:a3_configs_break}
     \vspace{-0.2in}
\end{figure}

\begin{figure}[t!]
     \begin{subfigure}[b]{0.4\columnwidth}
        \includegraphics[width=\columnwidth]{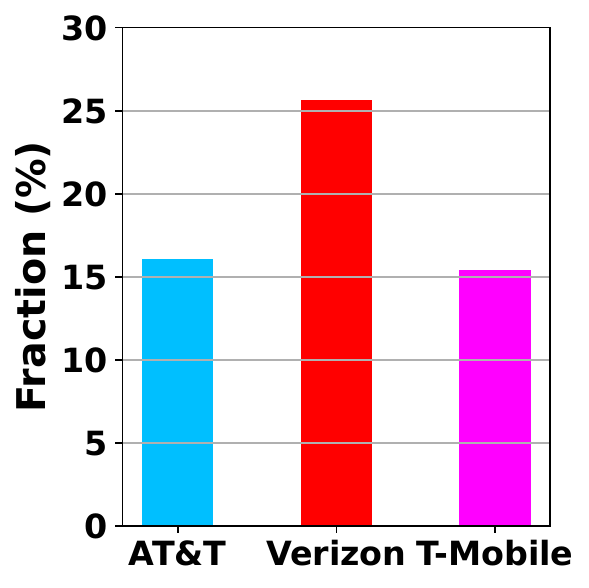}
        \vspace{-0.2in}
        \caption{Across operators.}
        \label{fig:a3_ping_pong_distribution}
    \end{subfigure}   
     \begin{subfigure}[b]{0.55\columnwidth}
        \includegraphics[width=\columnwidth]{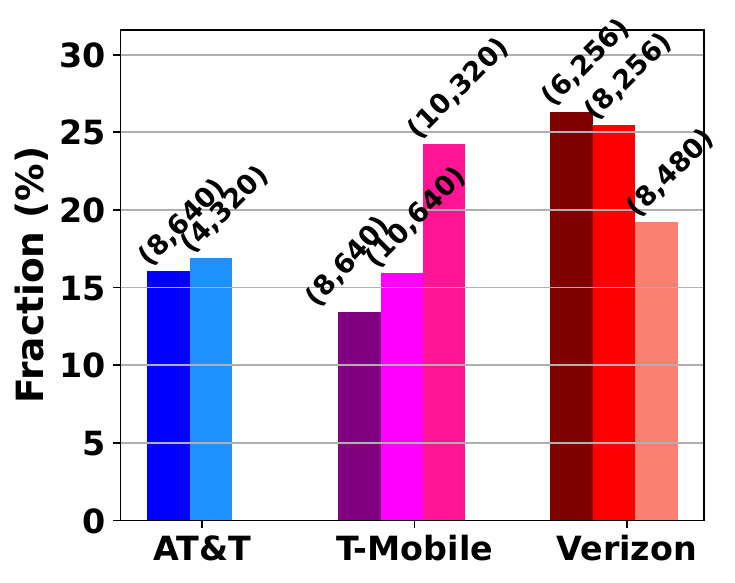}
        \vspace{-0.2in}
        \caption{Across A3 configurations.}
        \label{fig:a3_ping_pong_distribution_params}
    \end{subfigure} 
    \vspace{-0.1in}
    \caption{Fraction of A3 ping pong HOs.}
    \label{fig:a3_ping_pong_a3}
    \vspace{-0.2in}
\end{figure}

\noindent\textbf{RSRP analysis.} Fig.~\ref{fig:perf_intra_a3} plots $\Delta RSRP$ and $RSRP_{pre}$ CDFs for A3 HOs for all three operators. We observe that, regardless of the choice of the A3 parameters, 98\% or more of the A3 HOs result in a positive $\Delta RSRP$. This is a direct consequence of the condition of the A3 event, which triggers a HO when the neighboring cell's RSRP is higher than the serving cell's RSRP by an offset. However, the choice of
configurations has a direct impact on the relative performance
improvement among operators, as well as the performance
degradation before a HO, as we show next.

Fig.~\ref{tab:a3_intra} shows that AT\&T primarily uses one configuration (8, 640), while T-Mobile uses the same configuration 53\% of the time and another configuration with higher $\Delta_{A_3} + H_{A_3}$ (10, 640) 33\% of the time. In contrast, Verizon uses two configurations with lower TTT and/or offset 86\% of the time ((6, 256) and (8, 256)). As a result, both AT\&T and T-Mobile achieve similar $\Delta RSRP$, outperforming Verizon by 2 dB in the median case. However, the results for $RSRP_{pre}$ appear counter-intuitive. First, T-Mobile has much higher $RSRP_{pre}$ than AT\&T, although both operators use the same primary configuration. 
Second, Verizon exhibits the lowest $RSRP_{pre}$ among the three operators, even though it also achieves the worst $\Delta RSRP$. While Verizon's aggressive configurations explain the poor $\Delta RSRP$, they cannot explain the poor $RSRP_{pre}$.

This counter-intuitive result can be explained by the coverage differences among operators. 
In our dataset, T-Mobile has the highest 5G RSRP and Verizon
has the lowest (Fig.~\ref{fig:rsrp_dist}). This explains the higher $RSRP_{pre}$ for T-Mobile compared to AT\&T in Fig.~\ref{fig:perf_intra_a3} in spite of using the same (or a more conservative) configuration, as well as the lower $RSRP_{pre}$ for Verizon compared to the other two operators in spite of Verizon's more aggressive configurations.

 Figs.~\ref{fig:a3_param_rsrp_diff}, \ref{fig:a3_param_rsrp_before} further break down the impact of different configurations on $\Delta RSRP$ and $RSRP_{pre}$, respectively, for each operator, showing that \textit{the selected configurations do not always have the expected impact on performance}. In the case of AT\&T, the conservative configuration (8, 640) \textit{simultaneously optimizes both metrics}, achieving higher $\Delta RSRP$ than the more more aggressive configuration (4, 320), as expected, but also higher $RSRP_{pre}$, which is counter-intuitive. In the case of T-Mobile, the configuration with the small $TTT$ (10, 320) has the highest $RSRP_{pre}$, as expected, and lower $\Delta RSRP$ than (10, 640). However, between the two configurations with $TTT=640$, the one with the smaller offfset (8, 640) yields the best $\Delta RSRP$ 50\% of the time (better than (10, 640)), but the worst $\Delta RSRP$ 30\% of the time (worse than (10, 320)), while it also has the lowest $RSRP_{pre}$. In the case of Verizon, (8, 480) yields the best $\Delta RSRP$ and the worst $RSRP_{pre}$, as expected, due the much larger $TTT$. However, (8, 256) yields better $\Delta RSRP$ than (6, 256), as expected due to the larger offset, but also better $RSRP_{pre}$. 

To explore the root cause of these unexpected outcomes of the HO configurations, in Fig.~\ref{fig:a3_param_rsrp_neighbor} we plot the CDF of the RSRP of all the neighboring cells, measured by the serving cell. We observe that the relationship between the RSRP of neighboring cells for different configurations closely matches the relationship between the $RSRP_{pre}$ for the same configurations in Fig.~\ref{fig:a3_param_rsrp_before}. In the case of AT\&T, cells using the (4, 320) configuration are surrounded by neighbors with weaker signal strength compared to the neighbors of cells using the (8, 640) configuration. This suggests a sparser deployment in regions using the (4, 320) configuration compared to regions using the (8, 640) configuration, which probably explains the lower $RSRP_{pre}$ of the former. The same observation can be made for the T-Mobile configurations (8, 640) and (10, 640), and for the Verizon configurations (6, 256) and (8, 256).

\noindent\textbf{Ping pong analysis.} Fig.~\ref{fig:a3_ping_pong_distribution} shows that the fraction of A3 ping pong HOs is quite high for all three operators. As expected, Verizon experiences the highest fraction of ping pong HOs (25\% vs. 15-16\% for AT\&T and T-Mobile), owing to the low $TTT$ values (Fig.~\ref{fig:a3_tab_rsrp}), which result in signal strength measurements of low-confidence.
Fig.~\ref{fig:a3_ping_pong_distribution_params} shows the fraction of ping pong HOs across different configurations for each operator. For T-Mobile and Verizon, a longer $TTT$ (640 ms and 480 ms, respectively) results in a smaller fraction of ping pong HOs compared to a shorter shorter $TTT$ (320 ms and 256 ms, respectively), as expected. In contrast, for AT\&T, we observe a similar fraction of ping pong HOs with two very different configurations, (8, 640) and (4, 320), which suggests that, similar to signal strength, \textit{it is not always easy to predict the impact of an A3 configuration on ping pong HOs}.



\subsubsection{A5 HOs}
\label{sec:a5hos}

\begin{figure}[t]
	\centering
        \subfloat[Dominant combinations $((\Theta^S_{A_5} - H_{A_5}, \Theta^N_{A_5} + H_{A_5}), TTT)$.] {
\label{tab:a5_intra}
\resizebox{0.48\textwidth}{!}{%
\begin{tabular}{|c|c|c|}
\hline
\textbf{AT\&T}  & \textbf{T-Mobile} & \textbf{Verizon}  \\ \hline
\begin{tabular}[c]{@{}c@{}}((-31, -102), 640): 25\%\\ ((-104, -100), 320) : 16\%\\ ((-82, -110), 640): 16\%\\ ...\\  ((-122, -118), 160): 3\%\\ Count: 8\end{tabular} & \begin{tabular}[c]{@{}c@{}}((-101, -95), 320): 32\%\\ ((-115, -108), 640): 13\%\\ ((-31, -114), 640): 9\%\\ ...\\  ((-108, -108), 320): 2\%\\ Count: 14\end{tabular} & \begin{tabular}[c]{@{}c@{}}((-31, -103), 640): 33\%\\ ((-42, -118), 640): 10\%\\ ((-122, -111), 128): 8\%\\ ...\\ ((-115, -111), 640): 1\%\\ Count: 18\end{tabular} \\ \hline
\end{tabular}

}}\\
	\subfloat[Performance.] {
        \includegraphics[width=0.3\textwidth]{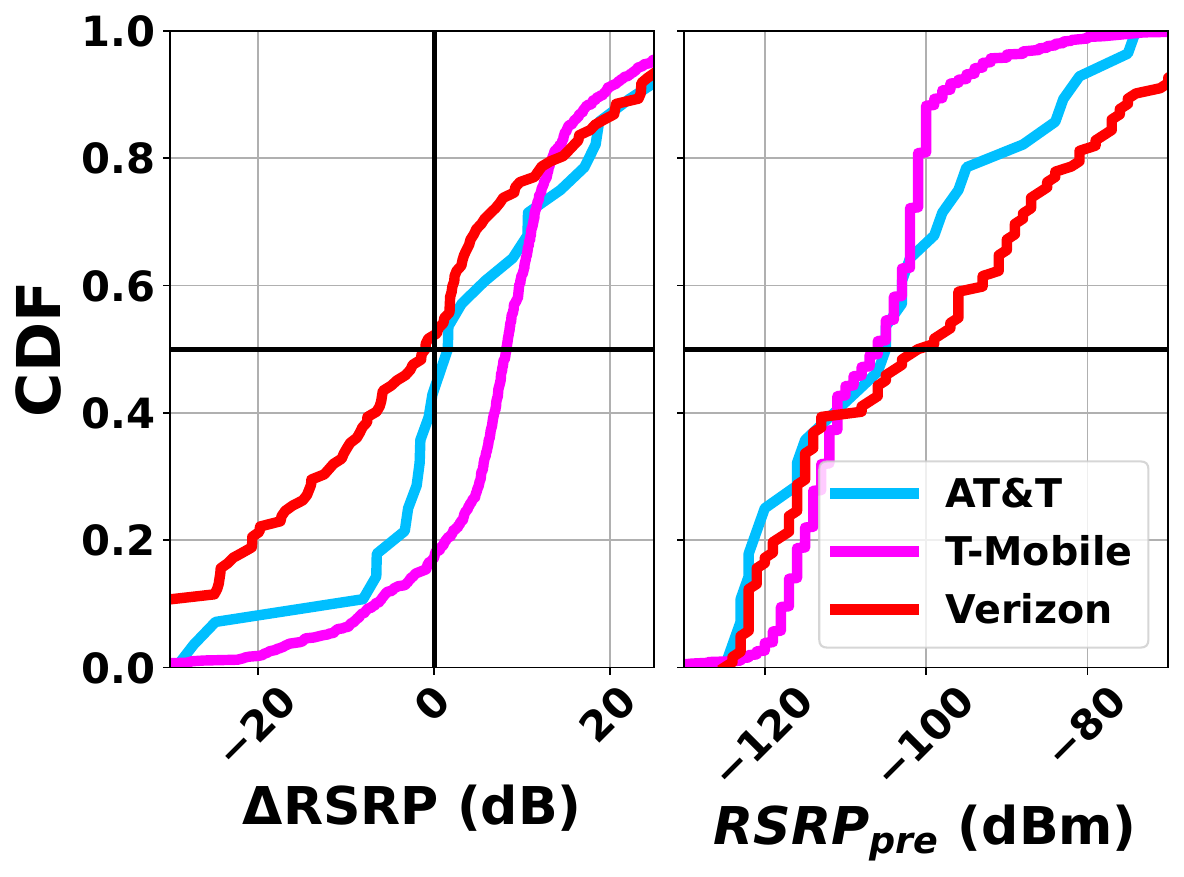}
        \label{fig:perf_intra_a5}
	}
	\subfloat[Ping pong HOs.] {
        \includegraphics[width=0.16\textwidth]{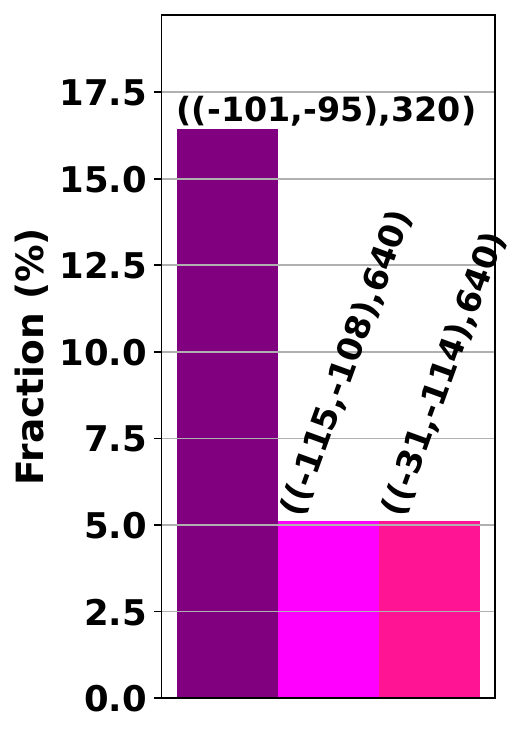}
        \label{fig:a5_ping_pong_distribution_params}
	}
	\vspace{-0.1in}
	\caption{Dominant parameter combinations and performance analysis for the A5 event.}
	\label{fig:a5_tab_rsrp}
    \vspace{-0.2in}
\end{figure}

An A5 event is triggered when two conditions are met for $TTT$ ms: the RSRP of the serving cell becomes lower than a threshold $\Theta^S_{A_5} - H_{A_5}$ and the RSRP of a neighboring cell becomes higher than another threshold $\Theta^N_{A_5} + H_{A_5}$. Fig.~\ref{tab:a5_intra} shows the dominant configurations -- tuples of $((\Theta^S_{A_5} - H_{A_5}, \Theta^N_{A_5} + H_{A_5}), TTT)$ -- for all three operators, which have a much higher richness than the A3 configurations. There are 8/14/18 dominant configurations for AT\&T/T-Mobile/Verizon, with no configuration appearing more than 33\% of the time. The configurations are again different across the three operators. 

\begin{figure}[t!]
    \subfloat[$\Delta_{RSRP}$.]{
        \includegraphics[width=0.48\columnwidth]{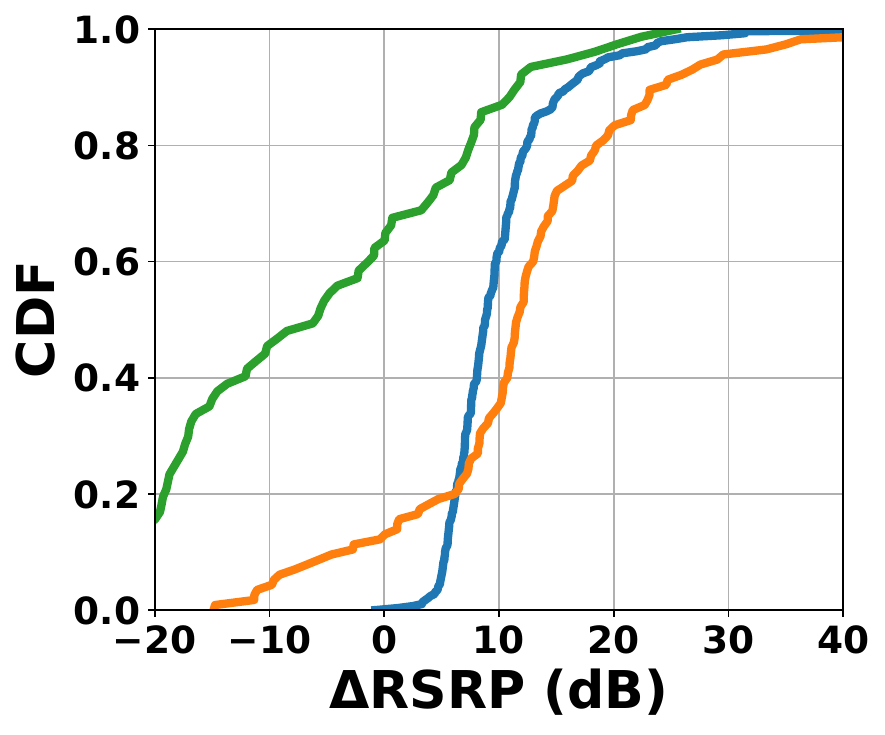}
        \label{fig:a5_param_rsrp_diff}
    }
    \subfloat[$RSRP_{pre}$.]{
        \includegraphics[width=0.48\columnwidth]{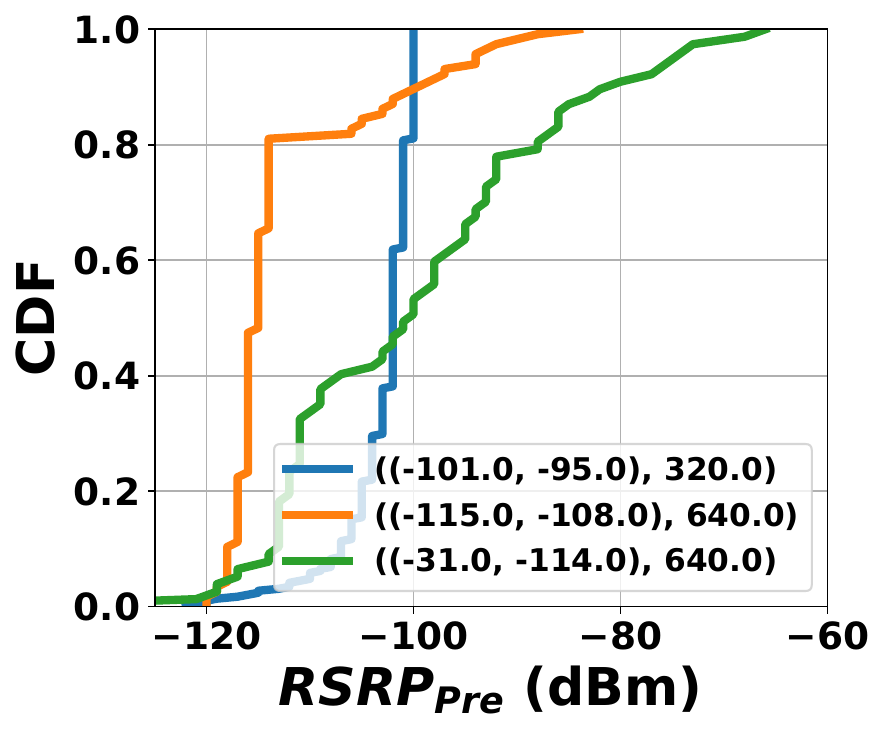}
        \label{fig:a5_param_rsrp_before}
    }
    \vspace{-0.1in}
    \caption{Impact of A5 configurations on performance.}
    \vspace{-0.3in}
\end{figure}

\noindent\textbf{RSRP analysis.} 
Fig.~\ref{tab:a5_intra} shows that the two primary configurations for Verizon -- ((-31, -103), 640) and (-42, -118), 640) -- and the primary configuration for AT\&T -- ((-31, -102), 640) -- are equivalent to A4 events, as $\Theta^S_{A_5} - H_{A_5}$ is very high (-31 or -42 dBm) and is almost always satisfied. Hence, an A5 MR is sent when a neighboring cell satisfies the second condition, regardless of the signal quality of the serving cell. As a result, these two configurations trigger many premature HOs that do not guarantee an improvement in signal strength, as shown in Fig.~\ref{fig:perf_intra_a5}; both operators have worse $\Delta RSRP$ than T-Mobile (negative about 50\% of the time) with Verizon achieving the worst performance. On the other hand, the same options ensure that the HO is often completed before the signal quality of the serving cell deteriorates significantly. This is reflected in Fig.~\ref{fig:perf_intra_a5}, which shows that Verizon has the highest $RSRP_{pre}$ among the three operators 60\% of the time, and AT\&T outperforms T-Mobile 35\% of the time. 

T-Mobile uses an A4-equivalent configuration ((-31, -114), 640) only 8\% of the time (vs. 25\%/43\% in the case of AT\&T/T-Mobile). As a result, it achieves the highest $\Delta RSRP$ among the three operators (positive 80\% of the time). It also achieves the highest $RSRP_{pre}$ 40\% of the time, primarily due to its aggressive $TTT$ in its primary configuration (320 ms vs. 640 ms) but lags behind Verizon 60\% of the time (and behind AT\&T 35\% of the time).

Figs.~\ref{fig:a5_param_rsrp_diff}, \ref{fig:a5_param_rsrp_before} further breaks down the impact of different configurations on $\Delta RSRP$ and $RSRP_{pre}$, respectively, for T-Mobile (we omit the other two operators, as their overall number of A5 HOs is very small -- 29 and 123 vs. 834). We focus on the top-3 dominant configurations for clarity. \textit{In contrast to the A3 event (Fig.~\ref{fig:a3_configs_break}), we observe that A5 configurations yield the expected outcome most of the time, as the use of two separate thresholds (compared to one offset in the case of A3) enable more fine-grained control}. The A4-equivalent configuration ((-31, -114), 640) yields the widest range of $RSRP_{pre}$ values and the lowest $\Delta RSRP$, as expected. Among the other 2 configurations, ((-115, -108), 640) results in a higher $\Delta RSRP$ (90\% of the time) and lower $RSRP_{pre}$ (70\% of the time) compared to ((-101, -95), 320) due to the larger $TTT$ and smaller $\Theta^S_{A_5} - H_{A_5}$. 


\noindent\textbf{Ping pong analysis.} Fig.~\ref{fig:a5_ping_pong_distribution_params} plots the fraction of A5 ping pong HOs for the top-3 configurations of T-Mobile. As expected, ((-101, -95), 320) results in a much higher fraction of ping pong HOs than than the other two configurations (16.3\% vs. 5\%) due to the much shorter $TTT$, in spite of the much higher $\Theta^N_{A_5} + H_{A_5}$ (-95 dB vs. -108/-114 dB) that should result in high 5G signal strength after the HO.

\begin{table}[t]
\centering
\caption{Dominant B1 combinations ($\Theta_{B_1} + H_{B_1}, TTT$).}
\label{tab:b1_inter}
\resizebox{0.4\textwidth}{!}{%
\begin{tabular}{|c|c|c|}
\hline
\textbf{AT\&T} &\textbf{T-Mobile} & \textbf{Verizon} \\ \hline
\begin{tabular}[c]{@{}c@{}}(-112, 80): 48\%\\ (-111, 40): 13\%\\ (-121, 80): 12\%\\ ...\\ (-117, 80): 4\%\\ Count: 6\end{tabular} & \begin{tabular}[c]{@{}c@{}}(-112, 80): 33\%\\ (-115, 80): 24\%\\ (-109, 256): 21\%\\ ...\\ (-103, 640): 5\%\\ Count: 5\end{tabular} & \begin{tabular}[c]{@{}c@{}}(-114, 256): 37\%\\ (-116, 256): 22\%\\ (-105, 256): 9\%\\ ...\\ (-111, 1280): 2\%\\ Count: 10\end{tabular} \\ \hline
\end{tabular}}
 \vspace{-0.1in}
\end{table}

\begin{figure}[t]
	\centering
        \includegraphics[width=0.45\textwidth]{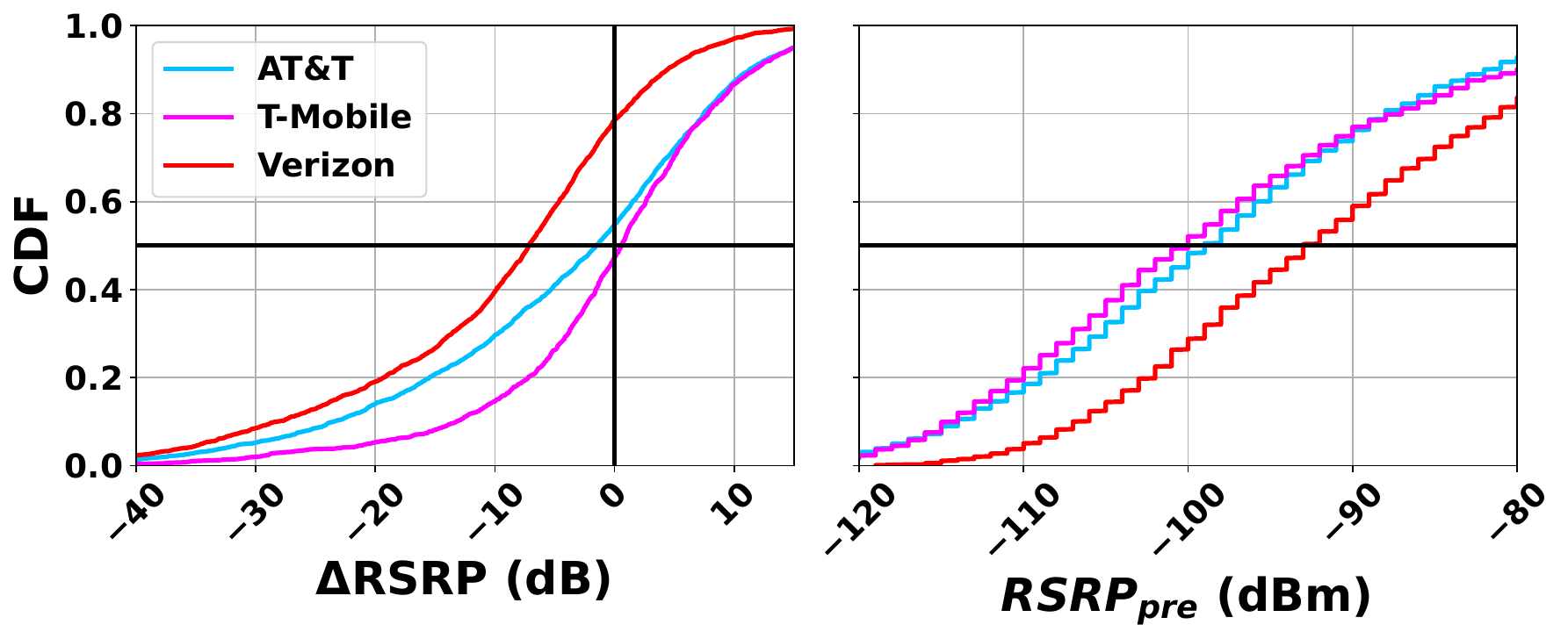}
	\vspace{-0.15in}
	\caption{RSRP analysis for B1 event.}
    \label{fig:perf_inter_b1}
     \vspace{-0.1in}
\end{figure}

\begin{figure}[t!]
     \begin{subfigure}[b]{0.4\linewidth}
        \includegraphics[width=\linewidth]{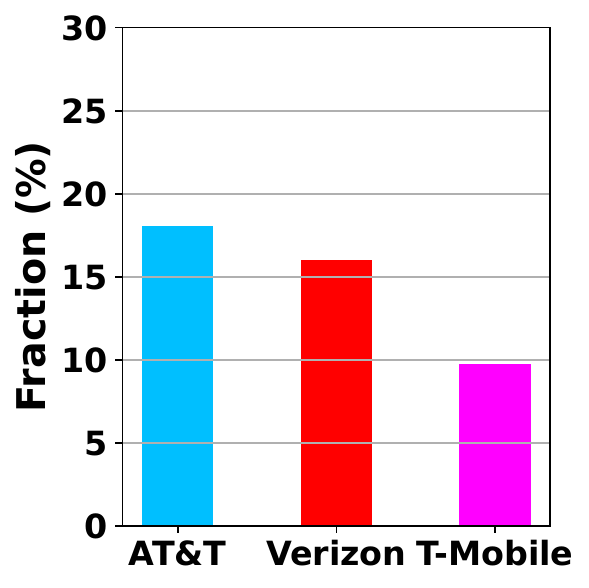}
        \vspace{-0.2in}
        \caption{Across operators.}
        \label{fig:b1_ping_pong_distribution}
    \end{subfigure}   
     \begin{subfigure}[b]{0.58\linewidth}
        \includegraphics[width=\linewidth]{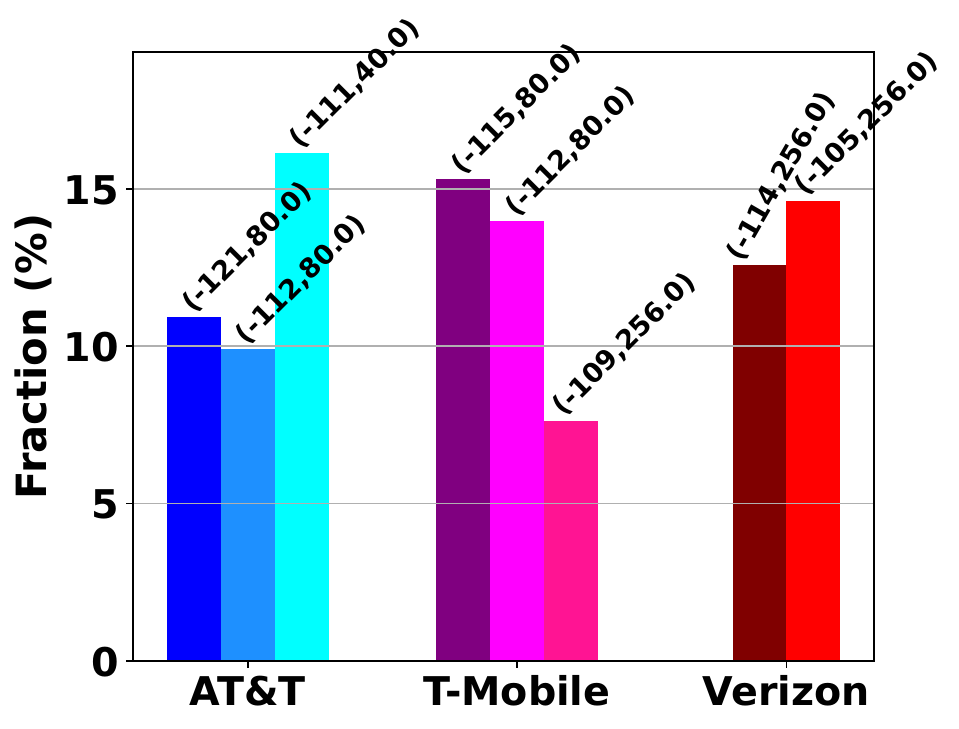}
        \vspace{-0.2in}
        \caption{Across B1 configurations.}
        \label{fig:b1_ping_pong_distribution_params}
    \end{subfigure} 
    \vspace{-0.1in}
    \caption{Fraction of LTE-5G ping pong HOs.}
    \label{fig:b1_ping_pong}
    \vspace{-0.2in}
\end{figure}

\subsubsection{B1 HOs}
\label{sec:b1hos}



B1 events are used to promote UEs from LTE to 5G. A B1 report is sent by the UE when the RSRP of an inter-RAT
neighboring cell remains higher than $\Theta_{B_1} + H_{B_1}$ for $TTT$. The B1 event does not take into account the RSRP of the serving (LTE) cell, as a promotion to 5G may be desired as long as the 5G cell's signal strength is satisfactory, independent of the serving LTE cell's coverage. 

Table~\ref{tab:b1_inter}
shows that AT\&T has 6 dominant configurations, T-Mobile has 5 dominant configurations,
and Verizon has 10 configurations. All three operators use similar values for $\Theta_{B_1} +
H_{B_1}$ but very different $TTT$ values. AT\&T uses the shortest
$TTT$ (40 ms and 80 ms) trying to aggressively promote UEs from LTE to
5G. T-Mobile uses both short and long $TTT$ values (80 ms, 256 ms, and
640 ms). Verizon, on the other hand, uses largely different TTT values for different 5G bands -- 128 ms (and rarely 1280 ms) for mmWave, 160-512 ms for 5G mid, 1024 ms and 1280 ms (and rarely 128 ms) for 5G low -- prioritizing HOs to 5G mmWave cells, which have limited coverage, and delaying promotion to 5G low.

\noindent\textbf{RSRP analysis.} 
Since the B1 event does not take the RSRP of the serving cell into account, an improvement in RSRP or a high $RSRP_{pre}$ are not guaranteed. This is clearly shown in Fig.~\ref{fig:perf_inter_b1},
where about 50\% of the B1-triggered HOs for AT\&T and T-Mobile and 80\% of
the B1-triggered HOS for Verizon lead to a reduction in the signal
strength (negative $\Delta RSRP$). The same figure shows that $RSRP_{pre}$ is typically higher than in the case of A3 or A5 HOs (median values  -100 to -93 dB for the three operators compared to -103 to -97 dB in Fig.~\ref{fig:perf_intra_a3} and -106 to -101 dB in \ref{fig:perf_intra_a5}). Fig.~\ref{fig:rsrp_dist} shows that Verizon overall has the highest LTE RSRP but the lowest 5G RSRP among the three operators, and its 5G RSRP is lower than its LTE RSRP. Consequently, Verizon has the highest $\Delta RSRP$ but the lowest $RSRP_{pre}$ in Fig.~\ref{fig:perf_inter_b1}. The high LTE RSRP but low 5G RSRP might also justify the long $TTT$ values used by Verizon, as the operator is trying to be conservative in upgrading a UE to 5G. On the other hand, T-Mobile has the highest 5G RSRP but the lowest LTE RSRP among the three operators (Fig.~\ref{fig:perf_inter_b1}). As a result, it achieves the highest $\Delta RSRP$ among the three operators in spite of using more aggressive TTT values than Verizon, but the lowest $RSRP_{pre}$ in spite of using more conservative TTT values than AT\&T.



\noindent\textbf{Ping pong analysis.} 
Fig.~\ref{fig:b1_ping_pong_distribution} plots the fraction of B1 ping pong HOs for each operator. T-Mobile has the lowest fraction of ping pong HOs (10\%) among the three operators, owing to its high 5G RSRP (Fig.~\ref{fig:rsrp_dist}). On the other hand, AT\&T aggressively promotes UEs to 5G using very short $TTT$ and experiences the largest fraction of ping pong HOs (18\%). Verizon uses longer $TTT$ than T-Mobile in its 3 primary configurations, but experiences a larger fraction of ping pong HOs (only slightly lower than AT\&T), due to its low 5G RSRP (Fig.~\ref{fig:rsrp_dist}).


Fig.~\ref{fig:b1_ping_pong_distribution_params} further breaks down the percentage of ping pong HOs across different configurations for each operator. We found the Verizon configuration (-116, 256) did not appear at all in 2024. Since the 2022 dataset contained a much smaller number of LTE-5G HOs and the 2023 dataset was collected only on the East Coast, Fig.~\ref{fig:b1_ping_pong_distribution_params} is based on the 2024 dataset only for a fair comparison across configurations. In the case of AT\&T and T-Mobile, we observe that the configurations with longer $TTT$ (80 ms and 256 ms, respectively) result in a much smaller fraction of ping pong HOs compared to the configurations with shorter $TTT$ (40 ms and 80 ms, respectively), as expected. Between configurations with the same $TTT$, one would expect that a higher $\Theta_{B_1} + H_{B_1}$ would mitigate the ping pong effect, as it upgrades the UE to a 5G cell with a stronger signal. Nonetheless, Fig.~\ref{fig:b1_ping_pong_distribution_params} shows that this is true only for AT\&T, suggesting that overall the $TTT$ parameter plays a much more critical role in controlling the ping pong effect.

\subsubsection{Summary}

Overall, selecting optimal configurations remains very challenging for operators, as they try to strike a balance between better post-HO performance and timing (which can deteriorate pre-HO performance), while also mitigating the ping pong effect. 
Optimizing all three metrics simultaneously is extremely tough, as two of these metrics -- pre-HO performance and ping pong HOs -- depend on additional factors, apart from the operator configuration, such as the overall coverage. In our analysis, we found only two configurations that simultaneously optimize 2 out of 3 metrics across the whole dataset, but not consistently across all combinations of timezone and area type, and no configuration that simultaneously optimizes all 3 metrics. 

Furthermore, in contrast to~\cite{deng:imc2018}, which
  reported that HO configurations affect radio signal as expected for
  LTE HOs, we found that this is not always true for 5G HOs. A3 HOs
  typically guarantee better post-HO signal strength but it is hard to
  predict the relative pre-HO (and sometimes post-HO) signal strength
  across configurations. On the other hand, it is easier to predict
  the relative performance of different A5 configurations, as they use
  two different RSRP thresholds for the serving and neighboring cell,
  instead of a relative offset, but A5 HOs often result in poor
  post-HO performance.

\subsection{RSRP vs. throughput improvement}
\label{sec:throughput}

In \S\ref{sec:rsrp}, we saw that operators cannot always guarantee an
improvement in RSRP after a HO, especially in the case of
B1 and A5 HOs. However, if such HOs result in an improvement in
throughput, they are successful from a user perspective. Hence, in
this section, we 
explore the impact of HOs on throughput in
relationship with the RSRP improvement. 
We calculate the average throughput 1 s before ($T_{pre})$ and 1 s after ($T_{post}$) a HO and then define $\Delta T = T_{post} - T_{pre}$. We also repeated the analysis with other time intervals between 0.5 s and 5 s and the results are similar.

\begin{figure}[t]
    \includegraphics[width=\columnwidth]{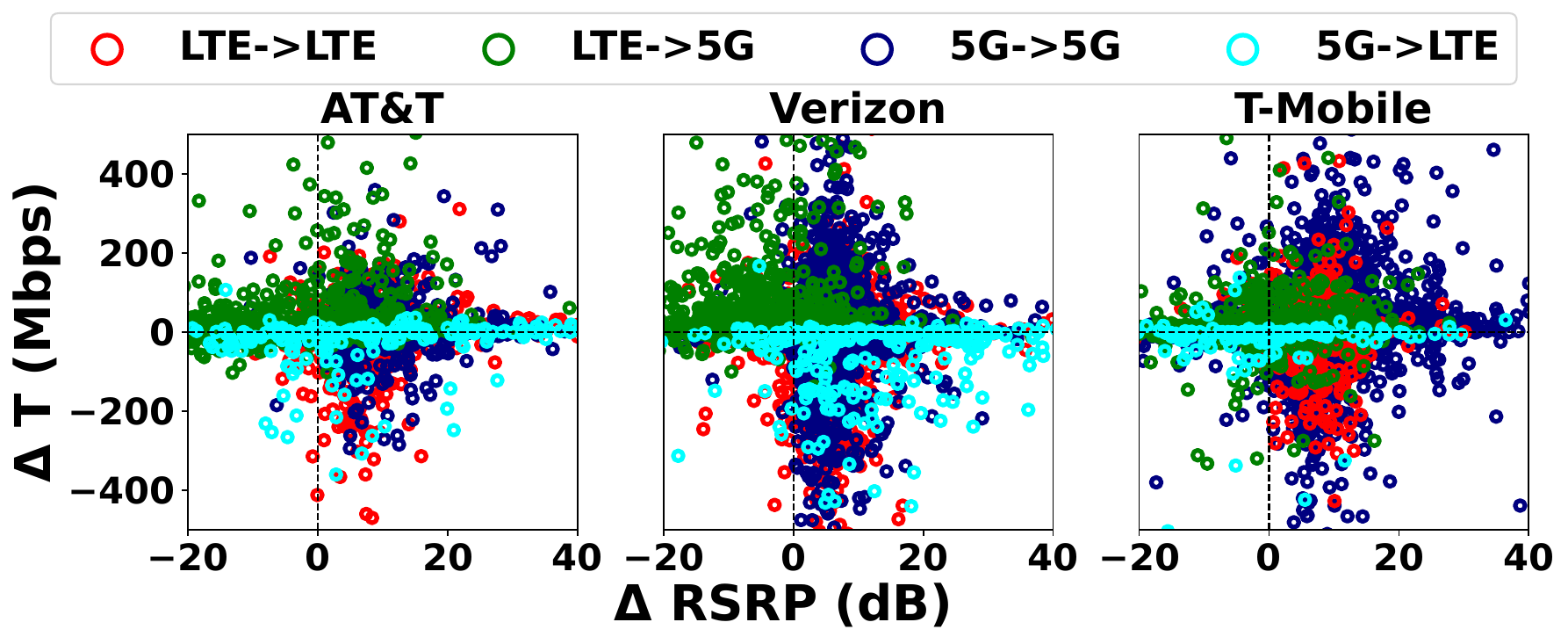}
    \vspace{-0.3in}
    \caption{RSRP and throughput difference for different types of HOs.}
    \label{fig:xput_rsrp_diff}
    \vspace{-0.3in}
\end{figure}

Fig.~\ref{fig:xput_rsrp_diff} shows scatterplots of $\Delta T$ against
$\Delta RSRP$ for all three operators and HO
types. We make the following observations:
(1) Overall there is \textit{no correlation between the improvement in signal strength and the improvement in throughput}. In particular, a positive $\Delta RSRP$ (which
operators struggle to achieve via their HO configurations) often leads
to a negative $\Delta T$ and vice versa. These results suggest that it is extremely hard
to optimize HOs from the user's perspective.
(2) The majority of intra-RAT LTE\textrightarrow LTE or 5G\textrightarrow 5G HOs (mostly triggered by A3 events) result in an improvement in RSRP, but this improvement does not guarantee an improvement in throughput. In fact, for all three operators, $\Delta T$ is 
negative for 40-50\% of the intra-RAT HOs.
(3) In contrast, 60\%/75\%/45\% of LTE\textrightarrow 5G HOs with
AT\&T/Verizon/T-Mobile lead to a reduced signal strength post-HO,
but $\Delta T$ is negative only for about 20-30\% of the HOs for all
three operators. In other words, aggressively upgrading a UE from LTE
to 5G often has the desired effect from the user's perspective even if
it leads to a signal strength drop, due to the
much higher bandwidth of the 5G mid and high bands compared to
LTE.
(4) Finally, 5G\textrightarrow LTE HOs 
lead to an RSRP improvement 85\% of
the time in the case of Verizon, but result in throughput
degradation 80\% of the time, and the degradation is more than 150 Mbps for 18\% of the HOs. In contrast,
AT\&T and T-Mobile appears to have a better control over 5G-LTE HOs and achieve
a better balance between RSRP and throughput. 

Overall, \textit{the weak correlation between RSRP (the metric that operators have direct control over via their policies) and throughput (the metric of interest from a user's perspective) makes it extremely tough to control the latter or predict its outcome based on the RSRP outcome.}

\section{Implications for Operators}
\label{sec:discuss}


We make five suggestions for operators to check their HO configurations based on the findings of our study. First, they should check their configuration parameters for horizontal LTE\textrightarrow LTE and 5G\textrightarrow 5G HOs. Our results (\S\ref{sec:throughput}) showed that, although the majority of those HOs improve RSRP, about half of them result in throughput degradation. Second, they should continue the trend we observed for T-Mobile (Fig.~\ref{fig:breakdown_2022_2024}) of replacing A3 events with A5 events, as A5 events have more predictable outcomes in terms of both pre- and post-HO performance (\S\ref{sec:a5hos}). However, extra care is required in configuring A5 events, as not all A5 configurations guarantee an RSRP improvement, unlike A3 configurations. We also found several problematic A5 configurations for 2 operators that ignore the serving cell's signal strength resulting in premature HOs. Third, they should refine 5G\textrightarrow LTE configurations, which appear to be overly aggressive in downgrading UEs from 5G to LTE, resulting in a significant throughput loss (\S\ref{sec:throughput}) despite RSRP improvement. Additionally, they should pay attention to various types of 5G failures, which are often responsible for downgrading UEs to LTE (\S\ref{sec:breakdown}). Fourth, they should pay extra attention to ping pong HOs. We found several overly aggressive configurations that result in a large number of ping pong HOs (up to 25\% of the total HOs), both intra- and inter-RAT (Figs.~\ref{fig:a3_ping_pong_distribution_params}, \ref{fig:b1_ping_pong_distribution_params}). Fifth, they should revisit all configurations based on the tradeoff between responsiveness and overhead. Our results (\S\ref{sec:events}) showed that a very large number of events are currently ignored by the network, resulting in unnecessarily high signaling overhead.

\section{Related Work}
\label{sec:related_work}

\noindent\textbf{HOs in the pre-5G era.} Several works studied HOs in operational networks in the pre-5G era~\cite{peng:icccn2016,li:ton2017,li:sigcomm2018,ni:nsdi2023,wang:mobicom2019,zhang:ton2023,li:sigmetrics2016,xu:pam2019,li:hotmobile2016,deng:imc2018}, focusing on the impact on performance~\cite{xu:pam2019,zhang:ton2023,li:ton2017,li:sigcomm2018,ni:nsdi2023,wang:mobicom2019,zhang:ton2023}, proposing solutions to mitigate this impact~\cite{xu:pam2019,zhang:ton2023,ni:nsdi2023}, or uncovering instability or unreachability issues~\cite{li:hotmobile2016,li:sigmetrics2016,peng:icccn2016}. The work closest to ours is~\cite{deng:imc2018}, which performed a detailed characterization of LTE HO configurations via a large-scale measurement study. 
In contrast to~\cite{deng:imc2018}, our work performs the first study of HO configurations in the 5G era.

\noindent\textbf{HOs in the 5G era.} A large number of experimental studies have evaluated 5G HOs over the past few years~\cite{narayanan:www2020,narayanan2021variegated,hassan:sigcomm2022,ghoshal:imc2023,xu:sigcomm:2020,khan:meditcom2024,kousias:Wintech2022,fiandrino:mswim2022,liu:infocom2023,zhang:conext2023,kalntis:imc2024,kalntis:infocom2025}. Most of these works~\cite{narayanan:www2020,narayanan2021variegated,xu:sigcomm:2020,khan:meditcom2024,kousias:Wintech2022,fiandrino:mswim2022,liu:infocom2023,zhang:conext2023,liu:mobicom2024} evaluate HOs on a small-scale, through mobility experiments spanning a few km or 1-3 of cities. 
The only works to our knowledge that examine 5G HOs on a large
scale are those in~\cite{ghoshal:imc2023,hassan:sigcomm2022,kalntis:imc2024}.
\cite{ghoshal:imc2023,hassan:sigcomm2022} use datasets from cross-country driving trips but focus on HO frequency, duration, and impact on transport and application layer performance. \cite{kalntis:imc2024} performs a country-wide analysis of HOs from an \textit{operator's perspective}, but uses a 4-week dataset and focuses on geo-temporal dynamics, frequency, duration, and causes of HO failures. Similar to~\cite{ghoshal:imc2023,hassan:sigcomm2022}, but different from~\cite{kalntis:imc2024}, our work studies HOs from the \textit{user's perspective}. Our work is the first to explore HO policies
and configurations in the 5G era, and therefore is complementary to~\cite{ghoshal:imc2023,hassan:sigcomm2022,kalntis:imc2024}. Further, our datasets span a 27-month period allowing us to study for the first time the evolution of 5G HO configurations.

\noindent\textbf{Diversity of cellular configurations.} Several works
(e.g.,~\cite{ge:mobicom2023,mahimkar:imc2022,mahimkar:sigcomm2021})
have studied cellular configurations 
and proposed data-driven or learning-based approaches for automated selection of parameter
values.
However, these studies are generic and not focused to HOs. Our work is
orthogonal to these works, focusing on HO configurations and comparing them across the three major US operators. 

\section{Conclusion}

In this work, we collected a large multi-carrier, multi-technology,
multi-band, cross-layer dataset of cellular mobility management during four cross-country driving trips spanning 15000+ km and spread out over a 27-month period. Leveraging the collected
dataset, we conducted the first in-depth
study of the HO landscape in the 5G era, analyzing different types of
HOs, including several new types that did not exist in the pre-5G era,
the configurations employed by different operators, the complexity of
those configurations, their diversity with respect to several
dimensions (technology, frequency band, 5G architecture, area type), their evolution, and their impact on performance. 
We expect that our findings and released dataset will
help both operators and the research community
to further
refine HO configurations and improve performance in next-generation
cellular networks.

\bibliography{reference}
\bibliographystyle{IEEEtran}


\end{document}